\documentclass[fleqn,useAMS,usenatbib,usegraphicx]{mn2e}
\usepackage{amssymb,amsmath}
\usepackage{mathrsfs}

\newcommand{\se}[1]{Section~\ref{sec:#1}}

\newcommand{\Se}[1]{\mbox{Section\ \ref{sec:#1}}}
\newcommand{\eq}[1]{equation~(\ref{eq:#1})}

\newcommand{\eqp}[1]{equation~\ref{eq:#1}}
\newcommand{\Eq}[1]{Equation~(\ref{eq:#1})}
\newcommand{\fg}[1]{Fig.~\ref{fig:#1}}

\newcommand{\eqs}[2]{equations\ (\ref{eq:#1}), (\ref{eq:#2})}

\newcommand{\Fg}[1]{Figure~\ref{fig:#1}}
\newcommand{\Tb}[1]{\mbox{Table\ \ref{tab:#1}}}
\newcommand{\app}[1]{\mbox{Appendix\ \ref{app:#1}}}
\newcommand\cf{cf.}
\newcommand\eg{e.g.,}
\newcommand\ie{i.e.,}
\newcommand\vs{vs.}

\newcommand\mEarth{\mathrm{M}_\oplus}
\newcommand{\pluto}{{\sc pluto}}

\newif\ifjournal
\journaltrue

\ifjournal

    \newcommand{\?}{}
    \newcommand{\refs}{}
    \newcommand{\gray}[1]{}
    \newcommand{\comq}[1]{}
    \newcommand{\com}[1]{}
    \newcommand{\combox}[1]{}
    \newcommand{\comno}[1]{}
    \newcommand{\ch}[1]{#1}
    \newcommand{\chk}{}

\else

    \usepackage{color}
    \definecolor{gray}{rgb}{0.5,0.5,0.5}
    \usepackage{framed}

    \newcommand{\chk}{\textcolor{red}{$^\textbf{[check!]}$}}
    \newcommand{\?}{\textcolor{red}{$^\textbf{?}$}}
    \newcommand{\refs}{\textcolor{blue}{$^\textrm{Refs}$}}
    \newcommand{\comq}[1]{\\ $\rightarrow$ \textcolor{red}{#1} \\}
    \newcommand{\com}[1]{\textcolor{blue}{[#1]}}
    \newcommand{\gray}[1]{\textcolor{gray}{[#1]}}
    \newcommand{\combox}[1]{\begin{framed}\textcolor{blue}{#1}\end{framed}}
    \newcommand{\comno}[1]{\begin{flushleft}\hspace{0.0\textwidth} \textcolor{blue}{Note:} \parbox[t][][t]{0.4\textwidth}{\small #1} \end{flushleft}}
    \newcommand{\ch}[1]{\textbf{#1}}

\fi

\title{Hydrodynamics of Embedded Planets' First Atmospheres. I. A Centrifugal Growth Barrier for 2D Flows.}
\author[Chris W. Ormel, Rolf Kuiper, Ji-Ming Shi]{
Chris W. Ormel$^{1}$\thanks{E-mail:ormel@astro.berkeley.edu}\thanks{Hubble Fellow},
Rolf Kuiper$^{2}$,
Ji-Ming Shi$^{1,3}$\\
$^{1}$Astronomy Department, University of California, Berkeley, CA 94720, USA\\
$^{2}$Max-Planck-Institut f\"ur Astronomie, K\"onigstuhl 17, D-69117 Heidelberg, Germany\\
$^{3}$Department of Astrophysical Sciences, Princeton University, Princeton, NJ
}
\begin{document}
\maketitle
\label{firstpage}
\begin{abstract}
    In the core accretion paradigm of planet formation, gas giants only form a massive atmosphere after their progenitors exceeded a threshold mass: the critical core mass. Most (exo)planets, being smaller and rock/ice-dominated, never crossed this line. Nevertheless, they were massive enough to attract substantial amounts of gas from the disc, while their atmospheres remained in pressure-equilibrium with the disc. Our goal is to characterise the hydrodynamical properties of the atmospheres of such embedded planets and their implication for their (long-term) evolution. In this paper -- the first in series -- we start to investigate the properties of an isothermal and inviscid flow past a small, embedded planet by conducting local, 2D hydrodynamical simulations. 
Using the \pluto\ code we confirm that the flow is steady and bound. This steady outcome is most apparent for the log-polar grid (with the grid spacing proportional to the distance from the planet).  For low-mass planets, Cartesian grids are somewhat less efficient as they have difficulty to follow the circular, large speeds in the deep atmosphere. Relating the amount of rotation to the gas fraction of the atmosphere, we find that more massive atmospheres rotate faster -- a finding consistent with Kelvin's circulation theorem. Rotation therefore limits the amount of gas that planets can acquire from the nebula. \ch{Dependent on the Toomre-Q parameter of the circumstellar disc, the planet's atmosphere will reach Keplerian rotation before self-gravity starts to become important.}
\end{abstract}


\section{Introduction}
\label{sec:intro}
Almost all planets carry an atmosphere. In the solar system rocky planets as the Earth, with a tiny atmosphere, can be distinguished from giant planets, where the atmosphere is substantial (Neptune) or where it dominates the mass of the planet (Jupiter).  Many exoplanets are likewise inferred to be rich in gas, even if they are of rather low mass \citep{MarcyEtal2014}. Of special interest to this work are the ubiquitous \textit{super-Earths} found by Kepler; their low bulk densities often betray a significant atmosphere with perhaps $1-10$\% of their mass in volatile elements \citep{LopezEtal2012}. Even those with a high density could have started with a significant atmosphere as in many cases the planet is so close to the star that the timescale for evaporation is short.

How do planets obtain an atmosphere? Tiny atmospheres --like Earth's-- can be obtained by outgassing \citep{SeagerDeming2010}. These are sometimes referred to as secondary atmospheres \citep{ZahnleEtal2007}.  More massive atmospheres can be captured from the gas of the circumstellar disc. This implies that the atmosphere formation is tied to planet formation -- \ie\ to the assembly of the planet's solid core and its interaction with the gaseous disc in which it resides for the first several million years. In order to capture any gas, planets have to be sufficiently massive. A first requirement is that the escape velocity from the planet's surface $v_\mathrm{esc}$ exceeds the magnitude of the thermal gas motions. Atmosphere formation therefore starts when:
\begin{equation}
  R_\mathrm{Bondi}
  \equiv \frac{G M_p}{c_s^2}
  \gtrsim R_\mathrm{surf},
  \label{eq:R-bondi}
\end{equation}
where $R_\mathrm{Bondi}$ will be referred to as the Bondi radius, $c_s$ is the isothermal sound speed, $R_\mathrm{surf}$ the radius of the solid core on top of which an atmosphere forms, $G$ is Newton's gravitational constant and $M_p$ the planet's mass, formally the sum of the mass of the solid core and the gaseous atmosphere, but at this stage fully determined by the mass of the core. 

\ch{A common definition for the atmosphere of the planet is the material within the Bondi radius, or, when it becomes large and tidal forces become important, the Hill radius, $R_\mathrm{Hill}=a_0(M_p/3M_\star)^{1/3}\gg R_\mathrm{surf}$. As \eq{R-bondi} scales linearly with mass, the outer radius quickly becomes much larger than the surface radius: the atmosphere is large and puffy. A key question is what the mass of this atmosphere is with respect to the total planet's mass (dominated by the core): $x_\mathrm{atm} = M_\mathrm{atm}/ M_\mathrm{p}.$ Initially, $x_\mathrm{atm}\ll1$ but $x_\mathrm{atm}$ increases as the planet accretes more solid material. This period is sometimes referred to as `Phase II'; it generally is a long phase as the liberated gravitational energy from the accretion counteracts the compression of the atmosphere \citep{PollackEtal1996,BodenheimerEtal2000,HelledEtal2013}. Alternatively, if impacting solids are absent, the atmosphere cools through Kelvin-Helmholtz contraction, likewise increasing $x_\mathrm{atm}$ but only very slowly for small planets \citep{IkomaGenda2006}. Once $x_\mathrm{atm}\approx1$ self gravity starts to become important and gas accretion strongly increases. The point where atmosphere and core mass are the same (crossover: $x_\mathrm{atm}=1/2$) is often taken as the definition of the critical core mass \citep[\eg][]{Rafikov2006}. }

\ch{When planets reach crossover, gas accretion accelerates \citep{BodenheimerPollack1986}; and once the disc can no longer supply the required amount of material, the atmosphere is no longer in pressure-equilibrium with the disc. }
Such high-mass planets may then accrete large amounts of gas from the disc. However, the observed amount of gas in giant planets is only a small fraction of the gas in the disc (a standard disc model typically has a mass $\approx$1\% of stellar, which is 10 Jupiter masses).  What has stopped a giant planet as Saturn from accreting more material is a key question that many groups are aiming to solve through state-of-the art hydrodynamical simulations \citep[\eg][]{LubowEtal1999,TanigawaWatanabe2002,Dobbs-DixonEtal2007,TanigawaEtal2012,GresselEtal2013,SzulagyiEtal2014}. Gap opening is one possibility; however, a very low viscosity is required to make a deep gap \citep{DuffellMacFadyen2013,FungEtal2014}.

This study does not concern this high mass regime. Rather, we will investigate the origin of primordial atmospheres around low-mass planets, which we define loosely as maintaining pressure-equilibrium with the disc \textit{and} having masses low enough to ensure the response of the disc to fall in the linear regime. The former condition requires that $x_\mathrm{atm}\ll1$ while the latter requires that $R_\mathrm{Bondi}<H$\refs.  We refer to the state of these atmospheres as \textit{embedded}.\footnote{\ch{\textit{Embedded} follows the terminology used in hydrodynamical simulations. Alternative terminology for an atmosphere in pressure-equilibrium with the disc are `nebular' \citep{BodenheimerEtal2000} or `attached' \citep{MordasiniEtal2012}.}} The embedded regime may apply to many super-Earths and to Neptune and Uranus, \ie\ for planets where the atmosphere is massive enough to guarantee a primordial origin, but where it does not dominate the total mass budget.
There are several questions regarding these embedded atmospheres:
\begin{enumerate}
  \item How much mass can the planet capture; what is the density and temperature structure?
  \item Is the atmosphere in steady-state?
  \item Is the atmosphere bound; that is, are streamlines closed and how does the atmosphere connect to the disc?
\end{enumerate}

The standard approach to model low-mass planet's atmosphere is by focusing on point (i), which requires us to solve the stellar structure equations of the atmosphere \citep[\eg][]{BodenheimerPollack1986,PapaloizouTerquem1999,Rafikov2006,MordasiniEtal2009,HoriIkoma2010,PisoYoudin2014}. Consequently, the 2nd and 3rd points are ignored: the atmosphere is assumed to be spherically symmetric, steady on a (hydro)dynamical timescale, and all gas within a certain radius (usually chosen the minimum of the Bondi and Hill radius, see \citealt{LissauerEtal2009}), is assumed bound. Assuming spherical symmetry, the atmosphere evolution is determined by the stellar structure equations. Nevertheless, these can become quite complex, as one must detail the properties of the gas: the equation of state (EOS), which will under high pressures deviate form ideal, the grain opacity, which can be altered by dust coagulation \citep{Podolak2003,MovshovitzEtal2010,Mordasini2014,Ormel2014}, and the heating due to solids falling into the atmosphere. Very crudely, the outcome of these calculations give a structure where the atmosphere, from top to bottom, is first isothermal with a temperature similar to the disc, then radiatively supported, and finally convective.

Despite the detail in these calculations, the question remains whether the neglect of points (ii,iii) is justified; indeed, whether hydrodynamic effects will not influence the thermodynamic properties of the atmosphere. However, to solve \textit{in addition} for the hydrodynamic structure is challenging. Mostly, hydrodynamic simulations concern the flow pattern at larger radii, away from the disc-to-atmosphere boundary, where they attempt to resolve the spiral density wave pattern that emerges \citep[\eg][]{PaardekooperPapaloizou2009,DongEtal2011}. Consequently, these studies ignore the detailed treatment of the atmosphere regions. A frequently-applied method is to consider vertically-averaged quantities by `softening' the gravitational potential by a length scale on the order of the disc scaleheight $H$. This has the advantage to speed up the simulations while still providing an accurate model for the torque density distribution the planes exert on the disc \citep[\eg][]{MuellerEtal2012}. However, such an approach also washes out any features on the scales of the atmosphere as, for low-mass planets, $H\gg R_\mathrm{Bondi}$.

In this and subsequent works we will conduct hydrodynamical simulations of a planet embedded in a gaseous disc with the prime aim of studying points (ii) and (iii) above. We compromise on point (i): the gas is for simplicity modeled as isothermal. Taking the thermodynamical properties of the gas into account via the energy equation is deferred to a further study. Here, the emphasis lies on the disc-atmosphere interface. In contrast to works studying the disc-planet interaction on larger scales, then, we must resolve scales well below the Bondi radius. Consequently, any softening of the gravitational potential of the planet must be limited to scales $\ll$$R_\mathrm{Bondi}$, implying a significant stratification of the gas. \ch{This particular combination of parameters---low planet mass ($R_\mathrm{Bondi}\ll H$), no accretion, and small softening lengths ($\ll$$R_\mathrm{Bondi}$)---makes for a challenging problem. It is an important phase to study, since all planets have once been small.}

This paper presents the result of local calculations conducted in 2D. which have the advantage of being computationally efficient. A key goal of this work is to verify whether the flow within the atmosphere is steady, as was assumed by previous semi-analytical work \citep{Ormel2013}. To facilitate the comparison, we introduce a polar grid \textit{centered on the planet} with a logarithmic spacing of the grid points. We find that this log-polar grid carries several advantages: (i) a natural refinement towards the areas characterised by steep gradients in density and velocity; (ii) a straightforward approach of handling the planet's surface as the inner boundary of the computational domain; and (iii) a remarkably good spatial resolution to computational expense ratio as compared to Cartesian geometries. We will conduct a parameter study to investigate the sensitivity of our results to variation of the key parameters (mass, surface radius). In particular, we will investigate the rotation profile of the atmospheres and detail under which conditions the atmosphere becomes rotationally supported.

This paper is structured as follows. In \se{setup} we present the governing equations, introduce the dimensionless units, the geometries, and the choice for the gravitational potential (softening parameter) of the planet. In \se{list} we present the simulation parameters. \Se{results} shows our results. We reflect on our findings in \se{discuss} and give our conclusions in \se{summary}.

\section{Setup of the hydrodynamical simulations }
\label{sec:setup}

\subsection{Governing equations, vortensity conservation}
We consider a compressible, inviscid, and isothermal fluid. The equations describing the flow are the continuity and Euler's equation:\footnote{The formulation here in this section uses densities $\rho$ as it applies generally (3D). However, the simulations are conducted in 2D for which density should be substituted by surface density $\Sigma$.}
\begin{align}
  \label{eq:continuum}
  \frac{\partial \rho}{\partial t} + \nabla \cdot \rho \mathbf{v} 
  &= 0 \\
  \left( \frac{\partial}{\partial t} +\mathbf{v}\cdot \nabla \right) \mathbf{v} 
  &= -\frac{\nabla P}{\rho} +\sum_i \mathbf{F_i}
  \label{eq:Euler}
\end{align}
where $\rho$ is density, $\mathbf{v}$ velocity, $t$ time, $P$ pressure, and $\mathbf{F}_i$ are externally-prescribed forces (accelerations). As the flow is isothermal, $P=\rho c_s^2$ where $c_s$ is the isothermal sound speed. \ch{We consider a local shearing flow with its reference frame comoving with the planet at the orbital frequency $\mathbf{\Omega}$. In this frame $\mathbf{x}$ points radially outwards (away from the star) and $\mathbf{y}$ in the (azimuthal) direction of the planet's orbit. The planet is assumed to move on a circular orbit}. The forces in this non-inertial frame read:
\begin{enumerate}
  \item The Coriolis force, $\mathbf{F}_\mathrm{cor} = 2\mathbf{\Omega}\times\mathbf{v}$, where $\mathbf{\Omega}$ is the local Keplerian frequency, which points in the $z$-direction, $\mathbf{\Omega} = \Omega \mathbf{e}_z$;
  \item The tidal force, $\mathbf{F}_\mathrm{tid} = 3x\Omega^2\mathbf{e}_x$ (with $\mathbf{e}_x$ the unit vector in the $x$-direction), which is the linearised contribution of the non-inertial centrifugal force and the solar gravity;
  \item The global pressure force due to sub-Keplerian motion of the gas, $\mathbf{F}_\mathrm{hw}=2\mathcal{M}_\mathrm{hw}c_s\Omega \mathbf{e}_x$, where $\mathcal{M}_\mathrm{hw}c_s$ is the lag in velocity with respect to Keplerian and $\mathcal{M}_\mathrm{hw}$ the Mach number of this headwind;
  \item The 2-body force due to the planet, $\mathbf{F}_\mathrm{2b}$, specified below.
\end{enumerate}
Self-gravity of the gas is ignored. In the case that the two-body force is absent, the unperturbed solution to the shearing-sheet reads:
\begin{equation}
  \mathbf{v}_\infty = \left(-\frac{3}{2}\Omega x -\mathcal{M}_\mathrm{hw}c_s\right)\mathbf{e}_y; \quad \rho=\rho_\mathrm{disc}
  \label{eq:v-inf}
\end{equation}
where $\rho_\mathrm{disc}$ is the gas density of the unperturbed disc.

Using some vector identities, the Euler \eq{Euler} can be rewritten (see \citealt{Ormel2013}):
\begin{equation}
  \frac{\partial \mathbf{v}}{\partial t}
  + \left( \mathbf{w} +2\mathbf{\Omega} \right) \times \mathbf{v}
  = -\nabla B 
  \label{eq:Euler-alt}
\end{equation}
where $\mathbf{w} = \nabla \times \mathbf{v}$ is the vorticity and $B=B(\mathbf{x})$ Bernouilli's `constant':
\begin{equation}
  B = 
  \frac{1}{2}v^2 
  +W 
  +\Phi_P 
  -\frac{3}{2}(\Omega x)^2 
  -2\mathcal{M}_\mathrm{hw} c_s x;
\end{equation}
and where it has been assumed that the flow is barotropic, such that the enthalpy $W$ satisfies $\nabla W = (\nabla P)/\rho$. The quantity $\mathbf{w}+2\mathbf{\Omega}$, sometimes called the absolute vorticity, is from now on denoted as $\mathbf{w}_a$.

We take the curl of \eq{Euler-alt} to find an equation for $\mathbf{w}_a$. Using standard vector identities
\begin{equation}
  \nabla \times (\mathbf{w}_a\times \mathbf{v})
  = \mathbf{w}_a (\nabla \cdot \mathbf{v}) +(\mathbf{v} \cdot \nabla)\mathbf{w}_a -(\mathbf{w}_a\cdot\nabla)\mathbf{v}
\end{equation}
and
\begin{equation}
  \nabla \times \nabla B = 0
\end{equation}
we rewrite
\begin{equation}
  \left( \frac{\partial}{\partial t} +\mathbf{v}\cdot \nabla \right)\mathbf{w}_a
  +\mathbf{w_a}(\nabla \cdot \mathbf{v})
  - (\mathbf{w_a} \cdot \nabla)\mathbf{v}
  = 0.
\end{equation}
Finally, divide this equation by $\rho$ and use the continuity equation to eliminate the $\nabla \cdot \mathbf{v}$ term. The first two terms can then be combined under one derivative, with the result:
\begin{equation}
  \frac{D }{Dt} \left( \frac{\mathbf{w}_a}{\rho} \right)
  =  \left[ \left( \frac{\mathbf{w}_a}{\rho} \right) \cdot \nabla \right] \mathbf{v}.
  \label{eq:zeta}
\end{equation}
For planar flow, \ie\ $\mathbf{v}=\mathbf{v}(x,y)$, $\mathbf{w}_a$ only has a $z$-component and the RHS of \eq{zeta} vanishes. This means that the quantity $w_{a,z}/\rho$ -- the \textit{vortensity} -- is conserved along streamlines. For the unperturbed flow, the vortensity is positive: $w_{a,z}/\rho=+\Omega/2\rho_\mathrm{disc}$ for an inviscid, barotropic flow in a Keplerian-rotating disc (note the plus sign: although the vorticity itself is negative [$-3\Omega/2$] the absolute vorticity $w_{a,z}$ becomes positive due to Coriolis effects). As the vortensity is spatially constant in the background and conserved along streamlines, it is the same $+\Omega/2\rho_\mathrm{disc}$ everywhere at all times, provided the flow stays isentropic (\ie\ does not shock). As the density is much higher the atmosphere (in the 2D approximation) is thus characterised by a large, and positive vorticity.

Previous works have used vortensity conservation to simplify the governing equations of motions \citep{KorycanskyPapaloizou1996,Ormel2013}. In this work we will instead use it to test the fidelity of the hydrodynamical code.


\subsection{Dimensionless quantities}
\label{sec:dimensions}
Let us from this point work in dimensionless units where velocities are expressed in terms of the sound speed $c_s$, times in terms of $\Omega^{-1}$, and lengths in terms of the disc scaleheight $H=c_s/\Omega$.  \ch{Dimensionless lengths are denoted by a lower case $r$: \eg\ $r_\mathrm{surf} = R_\mathrm{surf}/H$. In units of $c_s$ and $\Omega$, the planet's (gravitating) mass, denoted $m$, is defined as $m \equiv G M_p/(c_s^3/\Omega)$.} It follows that this is also the Bondi radius in dimensionless units: $m=R_\mathrm{Bondi}/H$. For comparison the dimensionless Hill sphere reads $r_\mathrm{Hill}\equiv R_\mathrm{Hill}/H = (m/3)^{1/3}$.

As we neglect self-gravity of the gas we have the liberty to set the background density unity; \ie\ gas densities are expressed in units of $\rho_\mathrm{disc}$, the (unperturbed) density of the circumstellar disc material at the position of the planet.  This means that the dimensionless gas mass of the atmosphere, $\mathscr{M}_\mathrm{atm}$, is in units of $\rho_\mathrm{disc} H^3$ (for 3D) or $\Sigma_\mathrm{disc} H^2$ (2D).

\Fg{dimQs} illustrates the link between physical and dimensionless quantities. The solid black lines give the physical mass of the planet as function of the position in the disc for three values of $m$ and for a disc temperature profile of
\begin{equation}
  T = 
  T_\mathrm{mmsn}
  = 270\ \left( \frac{a}{\mathrm{1 AU}} \right)^{-1/2},
  \label{eq:Tmmsn}
\end{equation}
(corresponding to the so called minimum-mass solar nebula model; \citealt{Weidenschilling1977i,HayashiEtal1985}).
The planet's physical mass is proportional to $m$ but it also strongly depends on the position of the disc: for a planet of a given physical mass, $m$ is lower in the outer disc.  The dimensionless mass also depends on the temperature in the disc: a hotter disc increases the disc scaleheight, causing the planet to becomes more embedded, \ie\ lower $m$. In \fg{dimQs} we have also drawn the surface radius of the planet normalised to the Bondi radius $R_\mathrm{surf}/R_\mathrm{Bondi}=r_\mathrm{surf}/m$, assuming a fixed internal density of $\rho_\bullet=5\ \mathrm{g\ cm^{-3}}$. A planet starts to bind gas when $r_\mathrm{surf}=m$ (\eqp{R-bondi}). For increasing planet mass and disc radius $r_\mathrm{surf}$ decreases.
\begin{figure}
  \includegraphics[width=88mm]{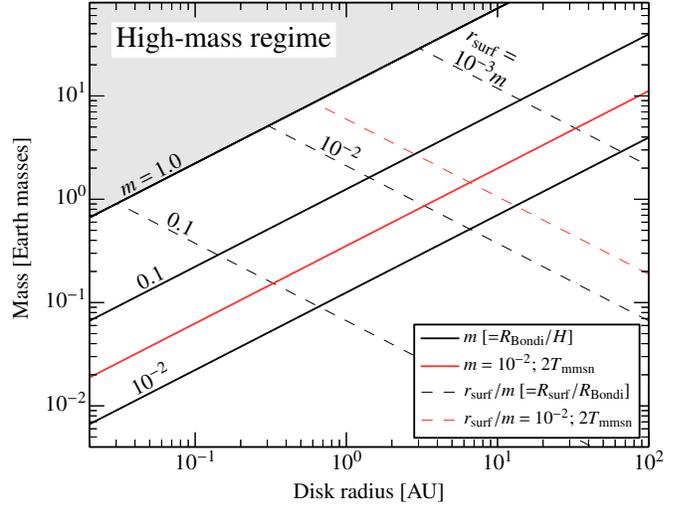}
  \caption{Correspondence between physical and dimensionless quantities. The axes denote the parameter space in terms of disc radius and planet mass. Contour lines of equal dimensionless mass $m$ (solid) and of equal planet surface radius-to-Bondi radius $R_\mathrm{surf}/R_\mathrm{Bondi}$ (dashed) are plotted. A solar mass star and a planet internal density of $5\ \mathrm{g\ cm}^{-3}$ have been assumed. Increasing the disc temperature shifts these lines upwards as indicated by the red lines for a disc that is twice the temperature of \eq{Tmmsn}. \ch{The high-mass regime ($m>1)$, shaded gray, is where the Bondi radius exceeds the Hill radius.}}
  \label{fig:dimQs}
\end{figure}

\subsection{Simulations geometries and boundary conditions}
We consider two geometries: Cartesian ($x,y$) and polar ($r,\phi$), as defined in 
\begin{eqnarray}
  x &=& r \cos \phi \\
  y &=& r \sin \phi
  \label{eq:xy}
\end{eqnarray}
Irrespective of the geometry, the planet is at the center of the coordinate system. The grid spacing of the Cartesian simulations is uniform, but in the polar runs we employ a logarithmic grid for the radial dimension. This has the advantage of concentrating grid points near the atmosphere region, where gradients in density and velocity are expected to be steep. In addition, the logarithmic spacing allows us to choose a large outer radius. We refer to the radially-logarithmic polar grid as \textit{log-polar}. 

We ensure to take the inner radius $r_\mathrm{inn}$ much smaller than the Bondi radius ($m$). The inner radial boundary is reflecting: \textit{there is no sink of gas}.  The boundary condition for the azimuth is periodic. In the Cartesian grid the $y$:$x$ aspect ratio is 2:1. Here, too, we take the unperturbed flow solutions (\eq{v-inf}) for the flow at the (outer) boundary; this is thus not a periodic boundary. In the Cartesian grid there is no inner domain edge: the two-body potential extends to $r=0$.  For both polar and Cartesian grid, the outer boundary conditions are taken, for simplicity, to be those of the unperturbed flow, \eq{v-inf}, \ch{as the outer boundary lies at a distance $\gg$$m$ (see \se{params}).}

\ch{We remark on a physical interpretation of the inner boundary in the log-polar runs $r_\mathrm{inn}$, which is not necessarily equal to the surface radius of the planet $r_\mathrm{surf}$. As we consider an isothermal EOS $r_\mathrm{inn}$ may be identified with the point where the atmosphere transitions from isothermal to optically thick or adiabatic. For smaller radii ($r<r_\mathrm{inn}$) the density profile will become a power-law \citep[\eg][]{Stevenson1982,Wuchterl1993,OrmelKobayashi2012,PisoYoudin2014}. Thus, $r_\mathrm{inn}$ can be found by calculating the $\tau=1$ optical depth layer (as measured from the surface), which depends on the density and the opacity of the gas. If the gas contains many small ISM-like grains, the opacity is high and the depth of the isothermal layer is modest. However, grain growth and grain settling can quickly remove the grains from these regions and---if these are not replenished---the opacity will be much smaller. For example, for a `virtually grain-free' 5 Earth mass planet at 5.2 AU ($m\approx0.1$) \citet{Ormel2014} calculates that the isothermal layer extends down to $\approx$$0.07$ of the Bondi radius.}

\subsection{The two-body force}
\begin{figure}
  \includegraphics[width=88mm]{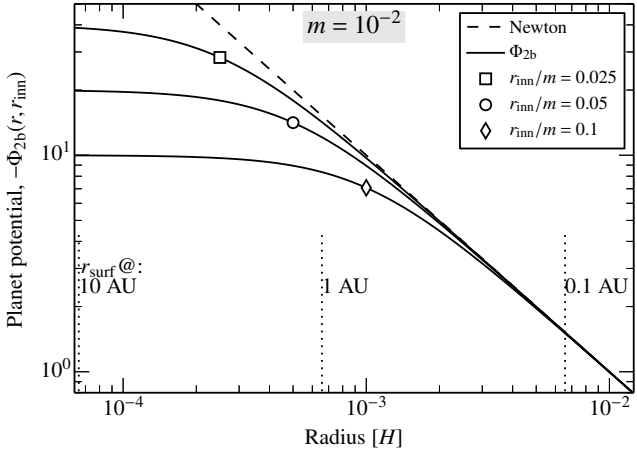}
  \caption{Comparison of the gravitational potentials that we use in this paper, $\Phi_\mathrm{2b}$ (\eqp{2-body}; solid curves), with the Newtonian $-m/r$ (dashed) for a dimensionless mass of $m=10^{-2}$. The potential is given for softening radii $r_\mathrm{inn}$ 2.5\%, 5\%, and 10\% of the Bondi radius. In the polar geometry $r_\mathrm{inn}$ also gives the start of the domain as indicated by the symbol. For Cartesian grids $r$ can be smaller than $r_\mathrm{inn}$. \ch{The vertical lines give the surface radius $r_\mathrm{surf}$ corresponding to several disc locations using the same parameters as in \fg{dimQs}.}}
  \label{fig:potentials}
\end{figure}
\begin{table}
  \centering
  \begin{tabular}{lp{57mm}}
  \hline
  \hline
  \multicolumn{2}{l}{Boundary conditions; unperturbed state} \\
  \hline
  \texttt{shear}          & static; \eq{v-inf} with $\mathcal{M}_\mathrm{hw}=0$  \\
  \texttt{headw}          & static; \eq{v-inf} with $\mathcal{M}_\mathrm{hw}=0.1$ \\
  \hline
  \multicolumn{2}{l}{Domain and geometry} \\
  \hline
  \texttt{Pol-extendD}    & polar, $(r,\phi)=[m/10:9]  \times[0:\pi]$ \\
  \texttt{Pol-ri20\%}     & polar, $(r,\phi)=[0.2m:0.5]\times[0:\pi]$\\
  \texttt{Pol}            & polar, $(r,\phi)=[m/10:0.5]\times[0:\pi]$ \\
  \texttt{Pol-ri5\%}      & polar, $(r,\phi)=[0.05m:0.5]\times[0:\pi]$\\
  \texttt{Pol-ri2.5\%}    & polar, $(r,\phi)=[0.025m:0.5]\times[0:\pi]$\\
  \texttt{Pol2pi}         & polar, $(r,\phi)=[m/10:0.5]\times[0:2\pi]$ \\
  \texttt{Cart2D}$^a$     & Cartesian, \newline $(x,y)=[-0.3:0.3]\times[-0.6:0.6]$ \\

  \hline
  \multicolumn{2}{l}{Masses} \\
  \hline
  (low)                   & low mass, $m=0.01$  \\
  \texttt{m0.03}          & intermediate mass, $m=0.03$ \\
  \texttt{m0.1}$^a$       & large mass, $m=0.1$ \\
  \hline
  \multicolumn{2}{l}{Resolutions$^{b,c}$} \\
  \hline
  \texttt{lowRs}          & 64x64 (polar);    256x512  (Cartesian) \\
  (medium)                & 128x128 (polar);  512x1024 (Cartesian)  \\
  \texttt{hiRs}         & 256x256 (polar);  1024x2048 (Cartesian) \\
  \texttt{ultRs}          & 512x512 (polar);  2048x4096 (Cartesian) \\
  \hline
  \multicolumn{2}{l}{Potential Injection times} \\
  \hline
  (default)               & $t_\mathrm{inj}=0.5$ \\
  \texttt{slow}           & $t_\mathrm{inj}>0.5$ \\
  \hline
  \hline
  \end{tabular}
  \caption{Resolutions at which the flow simulations are conducted. Note: symmetries may reduce the actual resolution used in the computation. Notes: 
  $^a$For the large mass runs the domain size and grid spacing are increased by a factor of two in the Cartesian runs; 
  $^b$The resolution in the radial direction increases in case of an extended domain while keeping the grid spacing constant;
  $^c$The resolution in the azimuthal direction of the polar grid runs reflect the full  $2\pi$. In the \texttt{shearPol} runs where only half of the domain is used the number of azimuthal grid points in the simulation is half of this number.}
  \label{tab:resolutions}
  \label{tab:list}
\end{table}
In dimensionless units, the Newtonian force reads $\mathbf{F}_\mathrm{N} = - \nabla \Phi_\mathrm{N} = -m/r^2\mathbf{e}_r$ with the potential $\Phi_\mathrm{N} = -m/r$ and $\mathbf{e}_r$ the unit vector in the radial direction. In the Cartesian geometry simulations, the singularity of the two body force at $r=0$ must be avoided by `softening' of the potential, \eg\ $\Phi_\mathrm{2b} = -m/\sqrt{r^2 +\epsilon^2}$ with $\epsilon$ the softening length. As explained in the Introduction a standard choice is to take $\epsilon\sim H$. But as our interest is rather on modeling the flow within the Bondi radius, we must choose a softening length $\ll$$R_\mathrm{Bondi}$ (or $\ll$$m$ in dimensionless units).


To enable a direct comparison, we will choose a $\Phi_\mathrm{2b}$ that is the same for the Cartesian and logpolar geometries:
\begin{equation}
  \Phi_\mathrm{2b}(r,r_\mathrm{inn}) 
  = - \frac{m}{\sqrt{r^2 +r_\mathrm{inn}^2}}.
  \label{eq:2-body}
\end{equation}
Thus, $\epsilon=r_\mathrm{inn}\ll m$ ensures that the softening length is small with respect to the Bondi radius. This potential is illustrated in \fg{potentials} for a planet mass of $m=10^{-2}$ and several choices for the inner radius $r_\mathrm{inn}$. Near $r_\mathrm{inn}$ the effect of softening is noticeable. The polar grid simulations have no grid points at radii less than $r_\mathrm{inn}$.

The default value of $r_\mathrm{inn}/m$ is 0.1. Potentially, the effects of (very) low $r_\mathrm{inn}$ can be dramatic as the solution for an isothermal, hydrostatic density profile of a pressure-supported atmosphere reads: 
\begin{equation}
  \rho 
  = \rho_\mathrm{hs}
  = \exp[-\Phi_\mathrm{2b}(r,r_\mathrm{inn})]
  \label{eq:rho-hs}
\end{equation}
From \fg{potentials} one sees that (for the Newtonian potential) peak densities will become calamitously large for $r\rightarrow0$, which would cause the atmosphere to collapse under its own weight. 

In reality, two effects prevent such a collapse for low-mass planets. First, it is unlikely that the atmosphere will be fully isothermal due to opacity of the gas (or rather the opacity of the dust grains in the gas). To solve in addition for the thermal state of the gas by including the energy transport equation is beyond the scope of this work. As explained above only the outer regions of atmospheres are isothermal and, hence, the inner domain or the softening radius $r_\mathrm{inn}$ reflects the point where isothermality ceases. 
Second, due to angular momentum conservation considerations the atmosphere gas spins up and provides rotational support. This effect is captured by our simulations and we will here investigate the transition from a pressure-supported to a rotationally-supported atmosphere.

In the simulations, the planet's potential is only gradually inserted; the planet's force evolves as:
\begin{equation}
  \mathbf{F}_\mathrm{2b} (\mathbf{x},t) 
  = -\nabla \Phi_\mathrm{2b} \left\{ 1 -\exp\left[ -\frac{1}{2} \left(\frac{t}{t_\mathrm{inj}}\right)^2 \right] \right\}.
  \label{eq:F2b-time}
\end{equation}
Our standard value for the injection timescale is $t_\mathrm{inj}=0.5$. That means that it is absent at the start of the simulations ($t=0$), whereas for $t>1$ it quickly reaches its asymptotic limit. Numerically, the rationale for inserting the planet gradually is to `cushion'\? the collapse of the gas, \ie\ to prevent shocks, as this will destroy the vortensity conservation. This is justifiable as the growth of the planet takes place on timescales much longer than $t_\mathrm{inj}$ -- too long, in fact, to follow with a hydrodynamical simulation.
\begin{figure*}
  \centering
  \includegraphics[width=180mm]{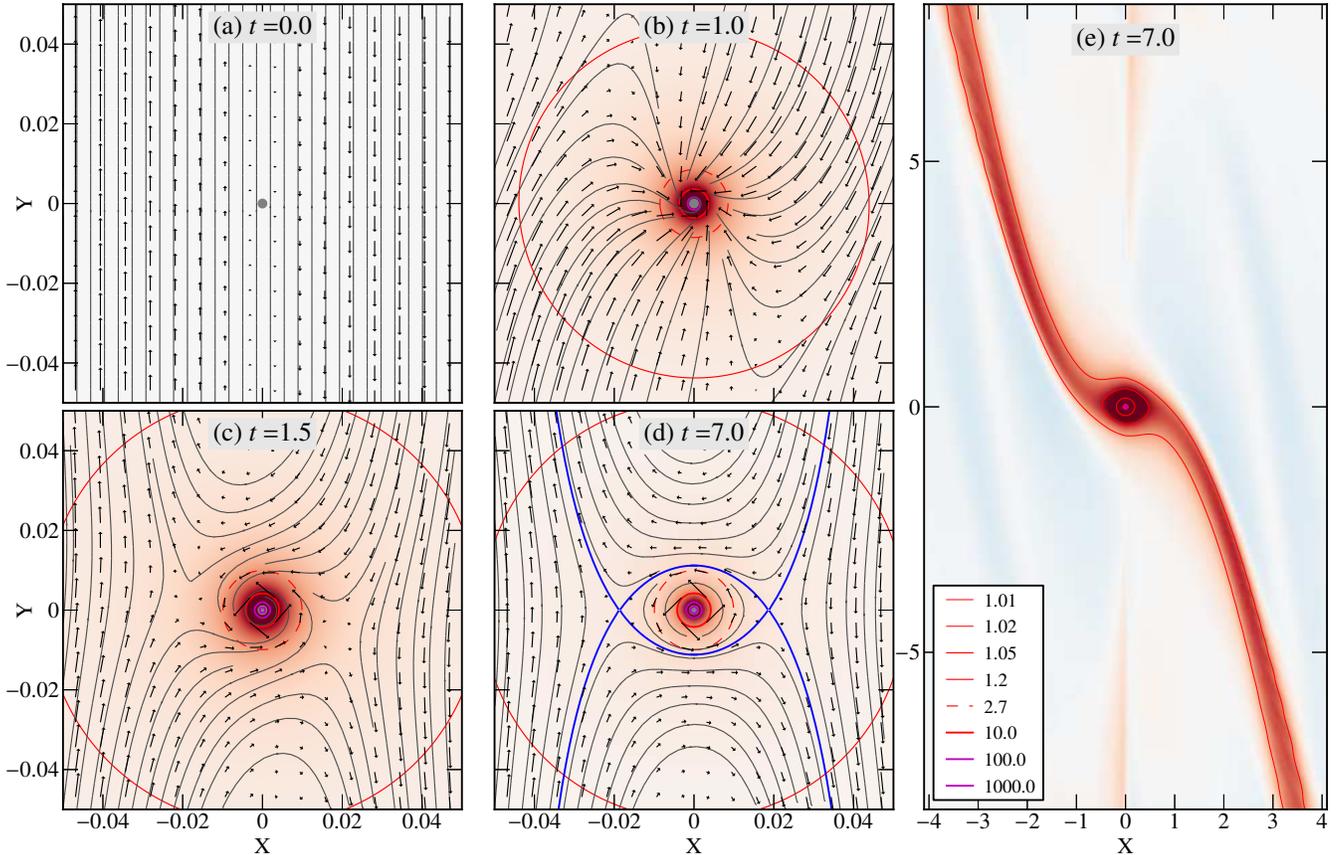}
  \caption{Flow pattern (density and velocity structure) of the \texttt{shearPol-largeD-hiRs} simulation run (2D, polar geometry, shearing sheet). The flow pattern is indicated by velocity vectors (arrows) and streamlines (gray curves). The density is indicated by the intensity of the orange/red background shade and by contour levels. (a) The unperturbed state (a shearing sheet); (b) the infall phase: the dashed red contour corresponds to $\rho=2.7$ and the large circle to $\rho=1.2$; (c) the transition to a pressure-dominated circulating atmosphere; (d) the final, steady state, where the planet atmosphere is isolated from the disc flow; (e) The large scale density wave structure. Note that the contrast level of the background differs per panel.  In panel (d) the thick blue curve gives the separatix streamline.} 
  \label{fig:collage}
\end{figure*}

\section{List of simulations}
\label{sec:list}

\subsection{The Pluto code}
We use the \pluto\ hydrodynamic code to solve \eqs{continuum}{Euler} \citep{MignoneEtal2007}. \pluto, version 4, is a publicly-available code, that includes state-of-the art algorithms for solving hydrodynamical and MHD flows. It is based on a high-order Godunov type shock-capturing Riemann solver. A variety of configurations\? are available for the numerical algorithm. We use fairly standard settings: a second order scheme in space and time (RK2), the Roe algorithm for our Riemann solver  and the Monotonized-Central flux limiter \citep{van-Leer1977}\com{Terminology OK?}.

One concern in choosing the solver is the effect of numerical viscosity.  In \app{sphere} we test the efficiency of the code by calculating the flow pattern of a uniform incident flow around a sphere, for which an analytic solution is available \citep{LandauLifshitz1959}. In the inviscid limit the flow velocity close to the surface of the sphere is larger than that of the unperturbed flow. However, numerical viscosity has a tendency to damp the flow motion, especially close to the sphere. We test two different solvers: the roe solver \citep{Toro1999} and the total variation diminishing flux Lax Friedrich (\texttt{tvdlf}) scheme \citep{TothOdstrcil1996}. We find that the Roe solver outperforms the tvdlf in terms of numerical viscosity. All our simulations in the main paper use the Roe solver.

\begin{table*}
  \centering
  \begin{tabular}{llrrrrrl}
  \hline
  \hline
  Name & $\langle \rho_X\ \rangle$       
             & rms $\rho_X$       
             & $\langle\langle w_z/\rho \rangle\rangle$   
             & rms $\langle w_z/\rho \rangle$   
             & $t_\mathrm{\rho X}$ 
             & $t_{wa/\rho}$
             & steady? \\
  \hline
  \texttt{shearPol-lowRs}              & {$8.9\times10^{2}$} & {$31$} & {$0.85$} & {$0.076$} & {$-4.1\times10^{1}$} & {$1.6\times10^{1}$} & no                                   \\
\texttt{shearPol}                    & {$1.0\times10^{3}$} & {$4.8$} & {$0.54$} & {$8.5\times10^{-3}$} & {$-3.2\times10^{2}$} & {$9.2\times10^{1}$} & no                                   \\
\texttt{shearPol2pi}                 & {$1.0\times10^{3}$} & {$3.8$} & {$0.54$} & {$8.2\times10^{-3}$} & {$-4.3\times10^{2}$} & {$9.5\times10^{1}$} & no                                   \\
\texttt{shearPol-hiRs}               & {$1.1\times10^{3}$} & {$0.55$} & {$0.50$} & {$8.1\times10^{-4}$} & {$-2.9\times10^{3}$} & {$9.0\times10^{2}$} & $\surd$                              \\
\texttt{shearPol-ultRs}              & {$1.1\times10^{3}$} & {$0.036$} & {$0.50$} & {$4.8\times10^{-5}$} & {$-4.8\times10^{4}$} & {$1.8\times10^{4}$} & $\surd$                              \\
\texttt{shearPol-m0.03}              & {$8.9\times10^{2}$} & {$4.5$} & {$0.52$} & {$4.1\times10^{-3}$} & {$-2.9\times10^{2}$} & {$1.9\times10^{2}$} & $\surd$                              \\
\texttt{shearPol-m0.03-hiRs}         & {$9.4\times10^{2}$} & {$0.77$} & {$0.50$} & {$4.8\times10^{-4}$} & {$-1.8\times10^{3}$} & {$3.4\times10^{3}$} & $\surd$                              \\
\texttt{shearPol-m0.1}               & {$5.2\times10^{2}$} & {$2.7$} & {$0.45$} & {$5.8\times10^{-3}$} & {$-2.8\times10^{2}$} & {$-1.8\times10^{2}$} & $\surd$                              \\
\texttt{shearPol-m0.1-hiRs}          & {$5.5\times10^{2}$} & {$0.24$} & {$0.44$} & {$7.1\times10^{-3}$} & {$-2.6\times10^{4}$} & {$-1.0\times10^{2}$} & $\surd$                              \\
\texttt{shearPol-largeD-hiRs}        & {$1.1\times10^{3}$} & {$0.79$} & {$0.50$} & {$7.7\times10^{-4}$} & {$-2.0\times10^{3}$} & {$9.5\times10^{2}$} & $\surd$                              \\
\texttt{shearPol-largeD-m0.1-hiRs}   & {$5.8\times10^{2}$} & {$0.35$} & {$0.50$} & {$4.8\times10^{-4}$} & {$3.5\times10^{3}$} & {$-1.9\times10^{3}$} & $\surd$                              \\
\texttt{shearPol-ri20\%}             & {$3.2\times10^{1}$} & {$0.034$} & {$0.19$} & {$0.049$} & {$-1.8\times10^{3}$} & {$-6.0\times10^{0}$} & no                                   \\
\texttt{shearPol-ri20\%-hiRs}        & {$3.3\times10^{1}$} & {$4.8\times10^{-3}$} & {$0.36$} & {$0.031$} & {$-1.2\times10^{4}$} & {$-1.8\times10^{1}$} & no                                   \\
\texttt{shearPol-ri5\%}              & {$4.3\times10^{4}$} & {$950$} & {$-1.8$} & {$1.2$} & {$-6.6\times10^{1}$} & {$2.5\times10^{0}$} & no                                   \\
\texttt{shearPol-ri5\%-hiRs}         & {$5.2\times10^{4}$} & {$410$} & {$-2.6$} & {$1.0\times10^{-3}$} & {$-1.8\times10^{2}$} & {$-3.6\times10^{3}$} & $\surd$                              \\
\texttt{shearPol-ri5\%-hiRs-slow}    & {$5.0\times10^{4}$} & {$290$} & {$0.48$} & {$2.2\times10^{-3}$} & {$-2.7\times10^{2}$} & {$-1.7\times10^{3}$} & $\surd$                              \\
\texttt{shearPol-ri2.5\%}            & {$3.0\times10^{5}$} & {$5.0\times10^{3}$} & {$-0.56$} & {$0.97$} & {$8.6\times10^{1}$} & {$-9.2\times10^{-1}$} & no                                   \\
\texttt{shearPol-ri2.5\%-hiRs}       & {$3.3\times10^{5}$} & {$6.4\times10^{3}$} & {$-2.1$} & {$0.17$} & {$-7.5\times10^{1}$} & {$-2.1\times10^{1}$} & no                                   \\
\texttt{shearPol-ri2.5\%-hiRs-slow}  & {$2.0\times10^{5}$} & {$5.5\times10^{4}$} & {$0.50$} & {$0.010$} & {$7.6\times10^{0}$} & {$-2.1\times10^{2}$} & no                                   \\
\texttt{headwPol2pi}                 & {$1.0\times10^{3}$} & {$3.6$} & {$0.47$} & {$6.7\times10^{-3}$} & {$-4.6\times10^{2}$} & {$-1.8\times10^{2}$} & $\surd$                              \\
\texttt{headwPol2pi-hiRs}            & {$1.1\times10^{3}$} & {$0.52$} & {$0.47$} & {$0.011$} & {$-3.2\times10^{3}$} & {$-1.4\times10^{2}$} & $\surd$                              \\
\texttt{shearCart2D-lowRs}           & {$2.0\times10^{0}$} & {$0.24$} & {$140$} & {$12$} & {$-2.0\times10^{1}$} & {$-1.2\times10^{2}$} & no                                   \\
\texttt{shearCart2D}                 & {$4.4\times10^{1}$} & {$4.2$} & {$-28$} & {$3.4$} & {$-1.2\times10^{2}$} & {$-3.0\times10^{2}$} & no                                   \\
\texttt{shearCart2D-hiRs}            & {$1.4\times10^{2}$} & {$2.5$} & {$-5.2$} & {$0.22$} & {$-1.2\times10^{2}$} & {$-3.5\times10^{1}$} & no                                   \\
\texttt{shearCart2D-ultRs}           & {$1.6\times10^{3}$} & {$0.37$} & {$-0.83$} & {$0.026$} & {$-6.5\times10^{3}$} & {$4.8\times10^{1}$} & no                                   \\
\texttt{shearCart2D-m0.03}           & {$7.4\times10^{2}$} & {$0.41$} & {$-0.058$} & {$5.2\times10^{-3}$} & {$5.9\times10^{3}$} & {$4.2\times10^{1}$} & no                                   \\
\texttt{shearCart2D-m0.03-hiRs}      & {$1.8\times10^{3}$} & {$4.5$} & {$0.81$} & {$0.030$} & {$5.8\times10^{2}$} & {$3.9\times10^{1}$} & no                                   \\
\texttt{shearCart2D-m0.1}            & {$8.2\times10^{2}$} & {$26$} & {$0.55$} & {$0.020$} & {$4.5\times10^{1}$} & {$7.1\times10^{1}$} & no                                   \\
\texttt{shearCart2D-m0.1-hiRs}       & {$9.0\times10^{2}$} & {$7.8$} & {$0.48$} & {$0.019$} & {$1.7\times10^{2}$} & {$2.3\times10^{2}$} & $\surd$                              \\

  \hline
  \hline
  \end{tabular}
  \caption{Output statistics of selected runs. Columns indicate: 
    run abbreviation (Name); 
    peak density averaged over at most 50 snapshots during $t=5$--10 ($\langle\rho_X\rangle$); 
    its root-mean square value (rms $\rho_X$); 
    atmosphere-averaged and time-averaged vortensity ($\langle\langle w_{z}/\rho\rangle\rangle$);
    its root-mean square value (rms $\langle w_a/\rho \rangle$); 
    depletion timescale (if negative) for the peak density obtained from a linear regression fit to $\rho_X(t)$ ($t_{\rho X}$); 
    same for the vortensity ($t_{wa/\rho}$); 
    assessment whether the run achieved steady-state or not (steady?).}
  \label{tab:outlist}
\end{table*}
\subsection{Parameters}
\label{sec:params}
A summary of simulation parameters is presented in \Tb{list}. The first column of of \Tb{list} gives the abbreviated name.  These abbreviations can be concatenated; for example, \texttt{shearPol} denotes a shearing sheet configuration ($\mathcal{M}_\mathrm{hw}=0$) in a polar (radially logarithmic) geometry at default values for the mass ($m=0.01$), resolution (128x128), inner radius 10\% of Bondi, and potential injection time. Not every parameter combination is tested.

Most simulations do not contain the headwind term (\texttt{shear}: $\mathcal{M}_\mathrm{hw}=0$). 
The advantage of the log-polar grid is that the outer radius can be easily extended to large values (\texttt{Pol-extend}). The inner radius' default is at 10\% of the Bondi radius ($r_\mathrm{inn}=0.1m$); but we will consider runs where $r_\mathrm{inn}$ in increased to 20\% of the Bondi radius (\texttt{Pol-ri20\%}) or decreased to 5 and 2.5\% of the Bondi radius (\texttt{Pol-ri5\%} and \texttt{Pol-ri2.5\%}). We consider three different dimensionless masses, $m=0.01$, 0.03 and 0.1 (\texttt{m0.01}, \texttt{m0.03}, \texttt{m0.1}), and four different resolutions -- low, medium, high and ultimate -- denoted by suffices \texttt{lowRs}, (none), \texttt{hiRs}, and \texttt{ultRs}, respectively. Generally, the non-uniform polar grid requires a smaller timestep than the uniform Cartesian grid. Therefore, the Cartesian grid has been assigned more grid points compared to the same resolution qualification of the polar grid. Our default injection time for the gravitational potential is $t_\mathrm{inj}=0.5$. However, in some cases this turns out to be too fast and we use an (arbitrarily) larger value. These simulations are denoted as \texttt{slow}.

For the \texttt{shearPol} runs we take advantage of the fact that the equations of motion employ a $\pi$-symmetry in polar angle; that is, $Q(r,\phi+\pi) = Q(r,\phi)$ for any quantity $Q$. Thus, we only simulate the upper plane ($0\le \phi \le \pi$) and use periodic boundary conditions for $\phi$. The resolutions listed in \Tb{list} always reflects the full $2\pi$ domain however. We have run one simulation with the full azimuthal domain for verification (\texttt{Pol2pi}).


\section{Results}
\label{sec:results}
\subsection{Emergence of a steady flow}
\Fg{collage} presents several snapshots at different times of the \texttt{shearPol-largeD-hiRs} simulation. Panels (a)--(d) show velocity arrows and streamlines. The panels' intensity\? reflects density, but the scaling varies from panel to panel. \Fg{collage}a shows the initial, unperturbed state at time $t=0$. The gas flow is that of the shearing sheet at unit density. At the very center of the image is the planet of radius $10^{-3}$, or 10\% of the Bondi radius, non-gravitating at this point.

\begin{figure}
  \includegraphics[width=88mm]{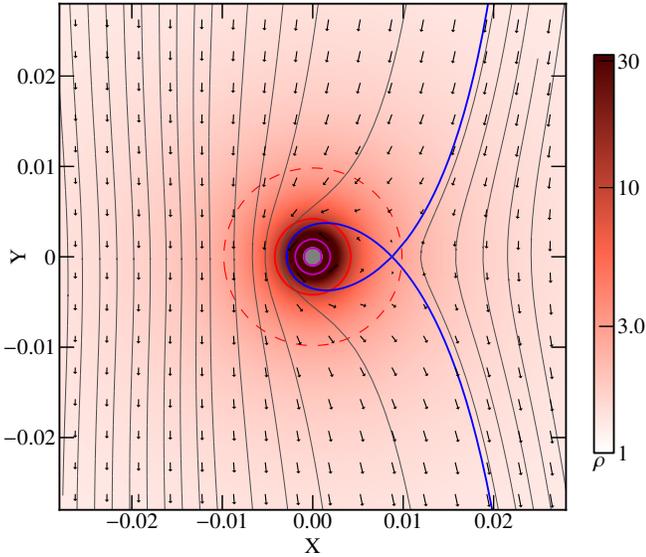}
  \caption{Flow within the atmosphere region of a planet that experiences a headwind (run \texttt{headwPol2pi-hiRs}).}
  \label{fig:headw-pol}
\end{figure}
The planet's gravity increases with time according to \eq{F2b-time}. By $t=1$ the potential has reached 86\% of its final value and is accreting copious amounts of gas. Remark that there is no mass flow of gas through the inner boundary: the gas just piles up near the center of the potential well. The infall is not completely radial: the flow is seen to spin up in the prograde direction: a consequence of the positive vortensity $w_a/\rho$ which is conserved in 2D.

After a time $t=1.5$ the planet's potential has reached 99\% of its final value. Material is still accreting but it is also evident that a disc has formed. Some of the incoming material is on horseshoe orbits; its orbit changes from interior to exterior (or vice versa) to the planet. (As our simulations are local we do not resolve the horseshoe region in its entirety; just the U-turn.)

Finally, for $t\gg1$ streamlines and velocity vectors no longer evolve. The flow is stable on timescales of $\sim$$\Omega^{-1}$, which we identify as steady. One recognises three kind of streamlines -- disc orbits (streamlines that orbit the star), horseshoe orbits (streamlines that make a U-turn) and circulating orbits around the planet. The streamline corresponding to the critical point where the velocity vanishes---the critical streamline (blue curve)---delineates these flow topologies.

These results are very robust: a steady state is reached at every resolution. In addition, the 2D runs are computationally efficient. The log-polar grid is an asset to resolve the flow within the Bondi radius as it concentrates grids at places where gradients in density and velocity are large. Nevertheless, the large scale features can still be reproduced. \Fg{collage}e shows the density pattern (streamlines are omitted as they all lie parallel on these scales) with its characteristic spiral wave. Note that the intensity levels of the density has been adjusted to highlight the wave, which is in reality overdense only at the 1\% level. At these scales, the resolution deteriorates and a number of minor artifacts are visible, most notably the density stripe at $X=0$. These are however numerical and their prominence decreases with increasing resolution.

\begin{figure}
  \centering
  \includegraphics[width=88mm]{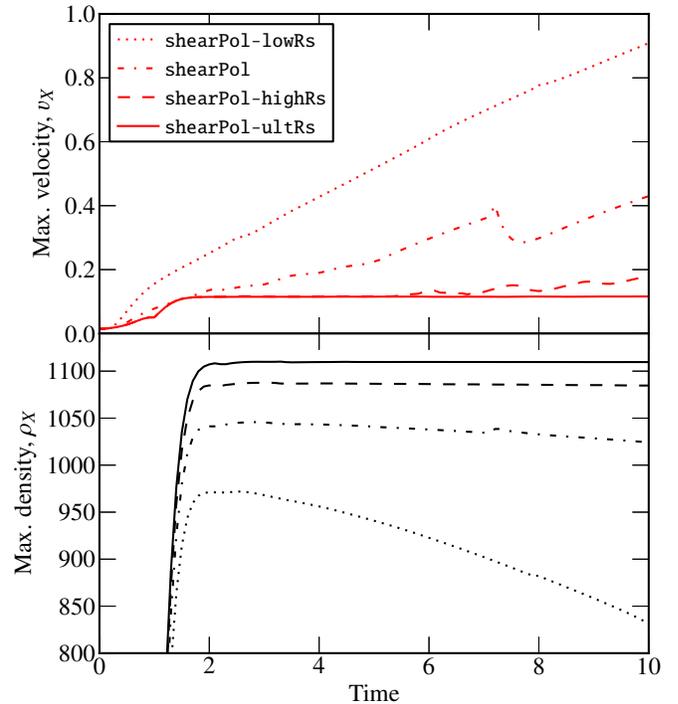}
  \caption{The maximum velocity (top panel) and maximum density (bottom) within the Bondi radius as function of time at several resolutions.}
  \label{fig:rhotime}
\end{figure}
\Fg{headw-pol} shows a zoom of the atmosphere region in case where a headwind term is present (\texttt{headwPol2pi}). The headwind term breaks the azimuthal symmetry. (Note that the apparent $Y$-symmetry breaks down at scales where the spiral density wave becomes apparent, see \fg{collage}e.) As a consequence of the headwind term the atmosphere becomes asymmetric. It is also much smaller; in this case it lies within the Bondi radius (the dashed red line in \fg{headw-pol}), whereas in the \texttt{shearPol} runs the atmosphere fully encloses the Bondi radius.

Qualitatively, these findings agree with \citet{Ormel2013}, where the steady-state form of \eqs{continuum}{Euler} are recast\? in terms of a convection-diffusion equation for the stream function (a scalar quantity), which is solved numerically.  The key advantage of the stream function approach is that it is quicker than running a hydrodynamical simulation. However, \pluto's 2D logpolar-grid simulations are computationally also very efficient; the moderate resolution is finished within an hour on a desktop PC. This suggests that one may as well opt to conduct direct hydrodynamical simulation.

\subsection{Quantifying the steady state}
\begin{figure*}
  \centering
  \includegraphics[width=180mm]{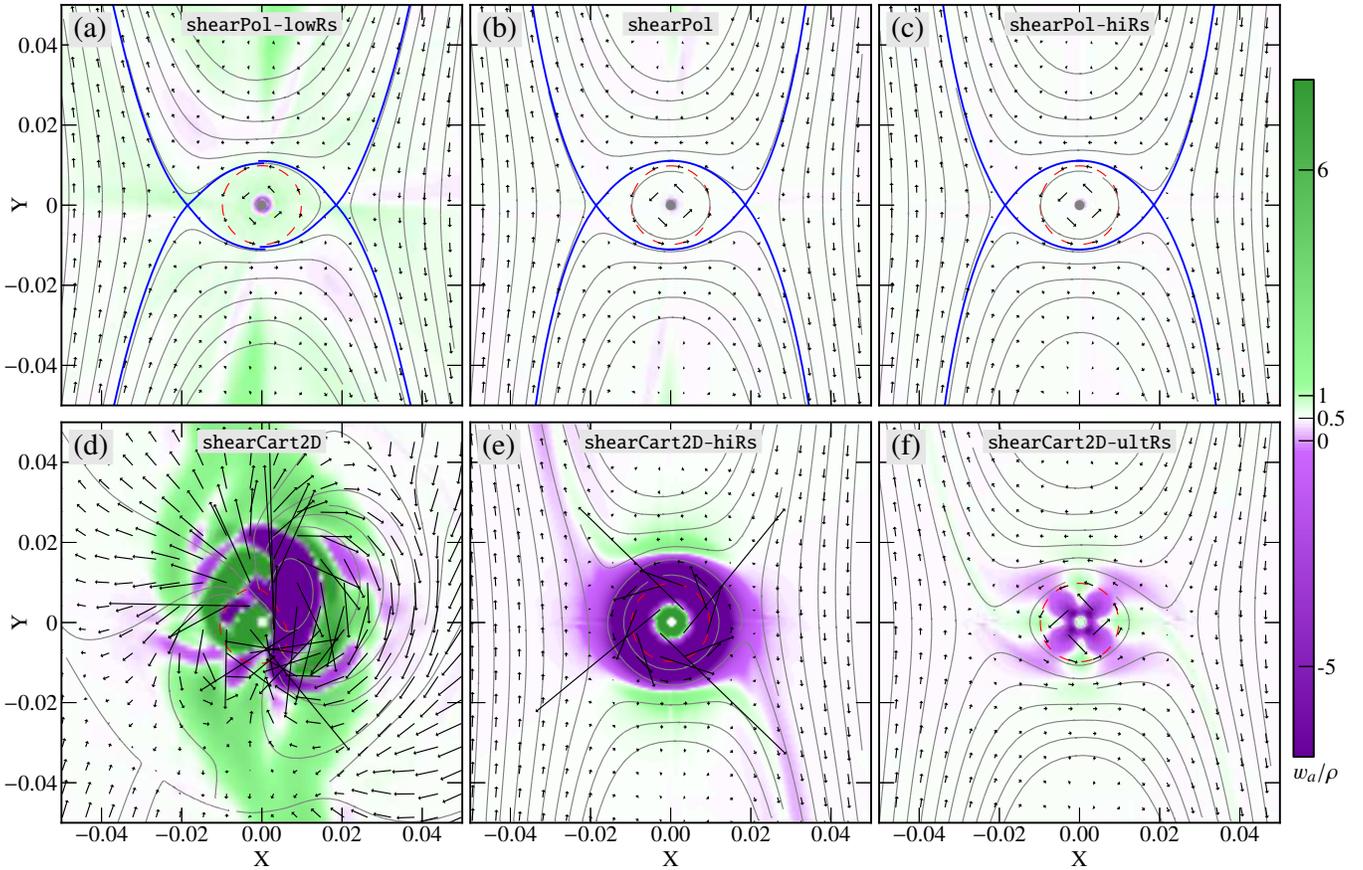}
  \caption{Resolution study for different geometries. The flow pattern is shown after $t=5$ for the log-polar (top) and Cartesian geometries (bottom) at increasing resolutions (left to right). The background color shows the vortensity $w_{a,z}/\rho$ of the flow. The polar runs reach a steady solution and the vortensity of the flow solution converges to the expected 0.5. The blue curves give the critical streamline. The Cartesian runs are non-steady and do not conserve vortensity, but a clear trend towards a steady outcome is seen with increasing resolution. The only density contour plotted is at $\rho=2.7$, which corresponds to the Bondi radius if the solution were static (red dashed).}
  \label{fig:geores}
\end{figure*}
\begin{figure*}
  \centering
  \includegraphics[width=180mm]{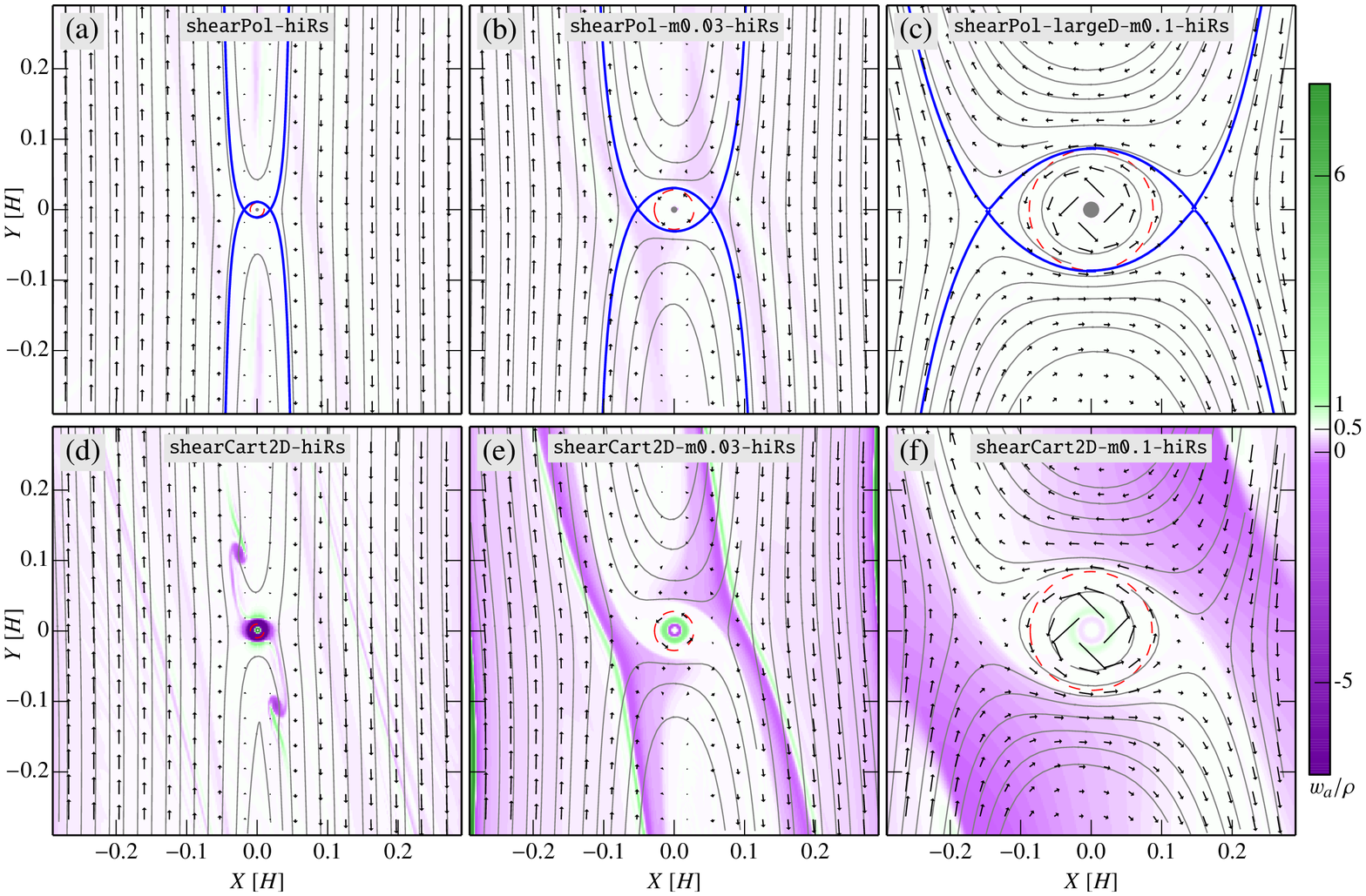}
  \caption{Effects of increasing planet masses (left to right: $m=10^{-2}$, $3\times10^{-2}$, and $0.1$). Top panels give the polar runs, bottom panels the Cartesian, all conducted at high resolution. The background color denotes the vortensity $w_{a,z}/\rho$. The diagonal stripes\? in vortensity seen especially in panels (b), (d), and (e) are boundary features that are advected with the Keplerian flow towards the atmosphere region. In panel (c) the domain of the simulation has been extended by a factor of 10, while keeping the log-grid spacing constant. In panel (f) both grid spacing and domain size have been increased by a factor two. The fidelity of the Cartesian simulations increases with increasing planet mass. \ch{The gray circles in panels (a)--(c) give the inner boundary of the domain.} }
  \label{fig:geomass}
\end{figure*}
Are the hydrodynamical simulations really in steady-state? In the previous section, our assessment of `steady flow' has been made mainly by eye, meaning for example that there are no clear jumps in density or velocity on short timescales. However, the system may none the less evolve on longer timescales.  

\Fg{rhotime} shows the maximum azimuthal velocity (upper panel) and the maximum density (lower panel) within the Bondi radius for the \texttt{shearPol} simulation at three different resolutions. Recall that the planet's potential is injected, so that at $t=0$ densities are $\approx$1. The peak density that is reached increases with resolution. This is simply because the first radial grid point lies closest to the inner radius ($r_\mathrm{inn}$) if the resolution increases. The relevance of \fg{rhotime} lies in the time-evolution of the curves.

The peak azimuthal velocities in the \texttt{shearPol-lowRs} run, for example, keep increasing with time and the densities decrease correspondingly. At $r=r_\mathrm{inn}$ the circular velocity corresponding to the potential of \eq{2-body} is $v_\mathrm{circ}=1.88$\chk. Thus, it seems that this simulations is progressing towards a rotationally-supported discs, at least where it concerns the very inner regions. At the same time the density is decreasing: material is leaking away from the atmosphere. This evolution all takes place close to the inner radius and hardly affects larger scales.

However, these trends are numerical as there is a clear resolution dependence. In the \texttt{shearPol} run the increase in velocity is less steep, and somewhat more erratic. Near $t=7$ the higher-rotating `mini-disc' near $r=r_\mathrm{inn}$ becomes unstable and the flow locally readjusts. The \texttt{shearPol-hiRs} shows very steady behavior in its peak density and only starts to increase its peak velocities after $t=5$. For the \texttt{shearPol-ultRs} no trends are apparent.

We attribute this resolution dependence to numerical viscosity, whose effects appear strongest towards the inner radius. On larger scales there is virtually no difference among the resolutions. The top panel of \fg{geores} plots the flow pattern and the vortensity $w_{a,z}/\rho$ after $t=5$ for the lowest three resolutions. Only the \texttt{shearPol-lowRs} shows clear deviations from the expected value +1/2 (corresponding to a white shading); the \texttt{shearPol} and \texttt{shearPol-hiRs} are virtually identical on these scales.

To get a measure on the steadiness of the simulation we have listed key statistics in \Tb{outlist} for a selection of runs. These show the behavior in peak density $\rho_X$ and atmosphere-averaged vortensity $\langle w_{a,z}/\rho \rangle$ in the $t=5$--10 time interval. We give the mean values of these quantities as well as their rms-deviations. We also give the trend of $\rho_X$ and $\langle w_{a,z}/\rho \rangle$ by applying a linear regression fit. The inverse of the slopes in these fit give the timescale on which the density and vortensity evolves: $t_\mathrm{\rho X}$ and $t_{wa/\rho}$.

From these indicators we determine whether the simulation run can be quantified as steady. For a steady solution we require that the timescale for the evolution of peak density and vortensity is much larger than the timescale over which the simulation is conducted (say by a factor 10). Thus, if any of the runs has a timescale that is $<$$10^2$ it is \textit{not} in steady state. In addition, we require that the rms fluctuations in $\rho_X$ and $\langle w_{a,z} /\rho\rangle$ are less than 5\%. The last column of \Tb{outlist} indicates whether the simulations obey these criteria.

Using these criteria we find that the polar grid simulations are generally steady, whereas the Cartesian simulations are not. Exceptions related to the arbitrary nature of the criteria are noteworthy. The \texttt{shearPol} and \texttt{shearPol2pi} simulations are quantified as unsteady as their $t_{wa/\rho}$ are just short of $10^2$. This is entirely due to the behavior of the flow near the inner edge of the domain (where we witness a speedup; \fg{rhotime}); on larger scales the flow of the \texttt{shearPol} \texttt{shearPol2pi} is stable. Note that the indicators of these two runs are also very similar, as they should be because of symmetry considerations.
The polar runs with smaller inner radii (\texttt{shearPol-ri5\%} and \texttt{shearPol-ri2.5\%}) are off in the averaged vortensity with respect to the expected value of $1/2$. We found that for these runs the initial \ch{accretion of the gas proceeded supersonically and shocks developed, violating vortensity conservation.} Adopting a larger $t_\mathrm{inj}$ restored the vortensity conservation \ch{as seen in the \texttt{shearPol-ri5\%-hiRs-slow} and \texttt{shearPol-ri2.5\%-hiRs-slow} runs. However, because of the longer $t_\mathrm{inj}$ the $r_\mathrm{inn}=2.5m$ run has not yet achieved a steady solution by $t=5$ and is thus (mis)identified as nonsteady.} 

\subsection{Dependence on geometry and mass}
\Fg{geores} shows the flow pattern of several shear-only simulations after a time $t=5$. The top panels show the log-polar runs, whereas the bottom panels show the (uniform) Cartesian runs. The resolution increases from left to right. The background plots the vortensity. Green colors denote high vortensity regions; purple are regions of low vortensity with respect to the background vortensity of $w_{a,z}=1/2$. The dashed red line corresponds to $\rho=2.7$, which approximately indicates the Bondi radius.

Between the polar grid runs the flow is qualitatively identical, with increasing resolution increasing the fidelity of the solution. Note the high vortensity regions near the $X=0$ line in the \texttt{shearPol-lowRs} run, which become larger away from the planet. These are numerical artifacts as these features disappear with increasing resolution. In the headwind runs ($\mathcal{M}_\mathrm{hw}\neq0$; not shown) these vortensity artifacts are more prominent. But these too diminish in amplitude with increasing resolution.

\begin{figure*}
  \centering
  \includegraphics[width=180mm]{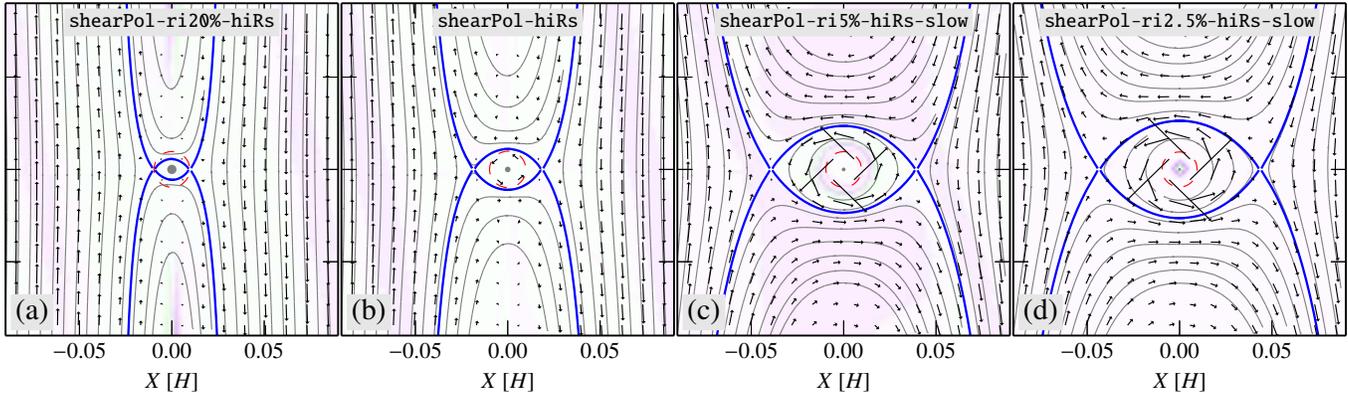}
  \caption{Flow pattern as function of decreasing inner radius: $r_\mathrm{inn}/m=0.2$ (left), 0.1, 0.05, and 0.025 (right). As $r_\mathrm{inn}$ decreases the size of the atmosphere expands, until the atmosphere reaches a point where it becomes rotationally supported. In the $r_\mathrm{inn}=0.05m$ and $r_\mathrm{inn}=0.025m$ runs the injection timescale $t_\mathrm{inj}$ was increased to 1.5 and 5.0, respectively, to ensure that the initial density pileup\? proceeds smoothly, preserving vortensity. Panels (a) and (b) show the flow pattern after $10t_\mathrm{inj}$ (or $t=5$); panels (c) and (d) after $5t_\mathrm{inj}$ ($t=7.5$ and 25, respectively).}
  \label{fig:rinns}
\end{figure*}
However, the outcome among the Cartesian runs is qualitatively different. The \texttt{shearCart2D} is very dynamic: the density and velocity of a grid point fluctuate widely. The \texttt{shearCart2D-hiRs} leads to a rotationally-supported disc of low peak density which `leaks' vortensity. The low vortensity feature is unsteady but wobbles greatly\?. The \texttt{shearCart2D-ultRs} run gives the closest match to the log-radial runs: the peak density is higher (the disc is pressure supported); the vortensity leakage is reduced with respect to the \texttt{shearCart2D-hiRs} run and decreases with time. The flow pattern is very similar to the polar run models. The only reason why the run was nevertheless quantified as unsteady was the lack of vortensity conservation within its atmosphere. 

The key explanation for the inferior performance of the \textit{uniform} Cartesian runs, is the lack of resolution in the region where it matters most: the atmosphere. For example the 512x1024 \texttt{shearCart2D} run has only 8.5 grid cells per Bondi radius whereas this number for the 64x64 \texttt{shearPol-lowRs} is 23.7. However, it also appears that the Cartesian runs have more difficulty with conserving vortensity in their atmospheres. We believe that the polar geometry is advantageous here because within the atmosphere the flow eventually becomes circular, aligned with the grid. On the other hand, in a Cartesian geometry circular motion is always misaligned by a certain amount: by moving the gas from one grid cell to the next, the distance to the planet changes slightly. This slight change in distance results in either the gravitational force to become larger than the pressure force (for radii larger than the equilibrium radius) or the pressure force to be larger than the gravitational (for radii smaller than the equilibrium).  As a result, the gas feels a force towards the equilibrium radius. Instead of moving force-free on a circular orbit (as for the polar geometry), the flow oscillates around its equilibrium radius, which causes it to interact, possibly adversely, with the flow at different radii. 

\Fg{geomass} plots the dependence of the flow pattern on planet mass, increasing it from the default $m=10^{-2}$ (left panels) to $3\times10^{-2}$ (middle) and $0.1$ (right). The log-polar runs all reach steady state. The size of the atmosphere clearly increases with mass (note that the range in $X$ and $Y$ is the same throughout the panels). In the \texttt{shearPol-m0.03-hiRs} run (\fg{geomass}b) one can see a patch of low-vortensity material. This is an artifact of the outer boundary (which lies at $r=0.5$), where we assigned the unperturbed quantities for $\rho$ and $\mathbf{v}$, discounting the influence of the planet's potential. To suppress these features, in \fg{geomass}c the outer boundary was moved outwards by over a factor 10, to $r\approx10$. Because of the logarithmic grid the increased computational expenses of this operation were minor.

In the uniform Cartesian runs we do not have the liberty to expand the domain boundary by such a jump without simultaneously sacrificing the resolution. The low-vortensity patches therefore are still prominent; for panel (d) and (e) $|X|=0.3$ is the boundary of the domain. In \fg{geomass}e one sees a large jump in vortensity near this boundary. In the \texttt{shearCart2D-m0.1-hiRs} we moved the outer boundary to $|X|=0.6$ but doubled the grid spacing, which, although somewhat ameliorating\? the boundary problem, is insufficient to resolve the spiral wave pattern (see \fg{collage}e).

Altogether, we see a marked increase towards the expected solution (steady flow and vortensity conservation) with increasing planet mass.  The \texttt{shearCart2D-m0.03-hiRs} gives a fairly steady flow, except for its vortensity, within its atmosphere.  The \texttt{shearCart2D-m0.1-hiRs} is the only Cartesian run to fulfill our steady-state criteria. In this run there are 170 grids/Bondi radius. Boundary effects, however, also become more problematic with increasing mass. Uniform Cartesian grids suffer either from an unresolved atmosphere (at low masses) or from an unresolved spiral wave pattern (at higher masses).


\subsection{The rotation profile}
\label{sec:rinn}
We test the sensitivity of the simulations towards variations in the inner radius $r_\mathrm{inn}$. Because of the isothermal EOS a decrease in $r_\mathrm{inn}$ implies a significant increase in density near $r_\mathrm{inn}$; if the atmosphere is pressure-supported it scales as $\rho\ \approx \exp(0.71 m/r_\mathrm{inn})$ at $r=r_\mathrm{inn}$ (\eqp{rho-hs}). In such isothermal models the bulk of the atmosphere mass resides near $r=r_\mathrm{inn}$. As $r_\mathrm{inn}$ decreases, more gas is thus accreted from the disc.

\Fg{rinns} shows the emerging flow pattern for the high resolutions. The \texttt{shearPol-ri5\%-hiRs} and \texttt{shearPol-ri2.5\%-hiRs} runs did not yield a steady flow, however (see \Tb{outlist}). The default injection timescale of $t_\mathrm{inj}=0.5$ is too short: the material falls in at a free-fall velocity and then shocks onto the atmosphere. This causes the atmosphere to become unstable. Therefore, the injection timescale of the potential was increased, by a factor of 1.5 and 5, respectively. As a result the infall was gentler (subsonic), conserving the vortensity. A stable disc formed. Note that even an injection timescale of $2.5$\chk\ (less than one orbital period) is much shorter than any conceivable assembly mechanism for the solid core: in reality the planet's core formed on a timescale much longer than the simulation times. This justifies the choice for a larger $t_\mathrm{inj}$ once we observe that the collapse proceeds too rapidly.

The key result of \fg{rinns} is that the size of the atmosphere (region of circulating flow) and the velocity within the atmosphere strongly increases with decreasing $r_\mathrm{inn}$.

\begin{figure}
  \centering
  \includegraphics[width=88mm]{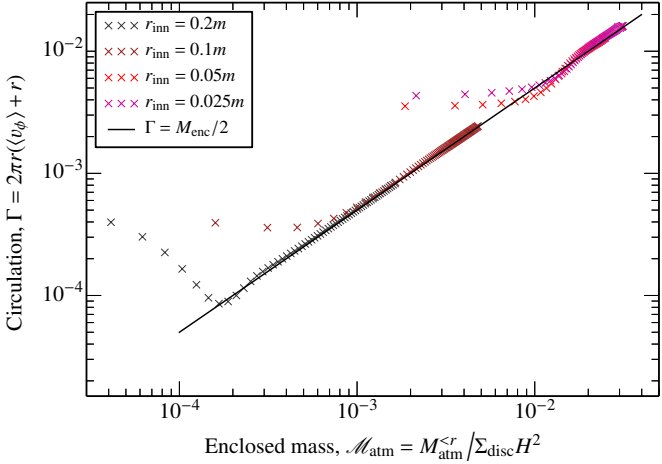}
  \caption{The amount of circulation, $\Gamma(r)$ (azimuthally-averaged) \vs\ the enclosed atmosphere mass $M_\mathrm{atm}^{r<}$ within radius $r$ at all grid points for the four runs of \fg{rinns}. The points collapse on a line $\Gamma = \mathscr{M}_\mathrm{atm}^{<r}/2$ consistent with Kelvin's circulation theorem.}
  \label{fig:atmos-circul}
\end{figure}
At first sight, this non-local effect is surprising: why should the velocity at a fixed point $r\gg r_\mathrm{inn}$ depend on the inner radius? This behaviour, however, follows directly from Kelvin's theorem on the conservation of circulation:
\begin{equation}
  \Gamma \equiv \oint (\mathbf{v} +\mathbf{\Omega} \times \mathbf{r}) \cdot d\mathbf{l} 
  = \int (\nabla \times \mathbf{v} +2\mathbf{\Omega}) \cdot d\mathbf{S}.
  \label{eq:circ}
\end{equation}
If we assume azimuthal symmetry and consider a closed contour of radius $r$ the first integral of \eq{circ} evaluates to $2\pi r(v_\phi +r)$. The second integral in \eq{circ} is a surface integration over the absolute vorticity $w_\mathrm{a,z}$. Here, we apply vortensity conservation, $w_\mathrm{a,z} = (w_\mathrm{a,z}/\rho)\rho = \rho/2$, and the integral simply evaluates to half of the enclosed mass within a radius $r$. Thus,
\footnote{\ch{The dimensional version of \eq{Kelvin-sol} reads
\begin{equation}
    2\pi R (v_\phi +\Omega R) = \frac{ M_\mathrm{atm}^{<r}\Omega }{2\Sigma_\mathrm{disc}}
    \label{eq:Kelvin-dim}
\end{equation}
}}
\begin{equation}
  \label{eq:Kelvin-sol}
  \Gamma 
  = 2\pi r (v_\phi +r)
  = \frac{\mathscr{M}_\mathrm{atm}^{<r}}{2}.
\end{equation}

In \fg{atmos-circul} we plot the predicted relation between $\Gamma$ and the enclosed mass. The simulation data are averaged over azimuth at every radial grid point starting from $r=r_\mathrm{inn}$ to $r=m$. The four different sets of curves correspond to the runs that differ by $r_\mathrm{inn}$, as in \fg{rinns}. Many points overlap and most of the points collapse on the predicted line (\eq{circ}). We attribute the offset seen for the inner few grid points to numerical (viscosity) effects, see \fg{rhotime}.


\begin{figure}
  \centering
  \includegraphics[width=88mm]{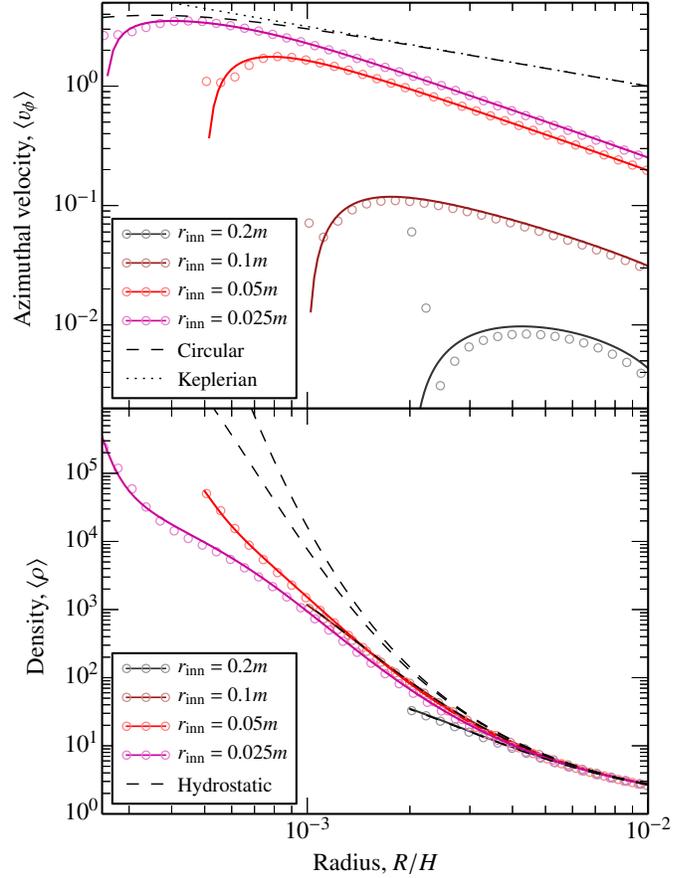}
  \caption{Azimuthal velocity (top) and density profiles (bottom) for three values of the inner boundary: $r_\mathrm{inn}=2.5\times10^{-4}$, $5\times10^{-4}$, $10^{-3}$, and $2\times10^{-2}$ (\ie\ 2.5\%, 5\%, 10\%, and 20\% of the Bondi radius). Circles denote the azimuthally-averaged simulation quantities while the solid curves give the solutions of \eqs{vort-simple}{Euler-simple}. In the top panel the dashed line gives the circular velocity and the dotted line the Keplerian. In the bottom panel the dashed lines give the hydrostatic solution \eq{rho-hs}.} 
  \label{fig:atmos-profile}
\end{figure}

The reason why the atmosphere becomes larger is because its mass increases. This is an effect (or artifact) of the isothermal EOS, which causes the mass of the atmosphere to be concentrated near the inner radius. Physically, when decreasing $r_\mathrm{inn}$ more material is `sucked' into the atmosphere; the material that now encircles the planet at radius $r$ originated from further out and was spun up. However, this mechanism can only operate when the atmosphere mass is dominated by the material near its inner radius. In \fg{rinns}d this stalls the growth of the atmosphere.

Assuming azimuthal symmetry and ignoring the Coriolis and tidal terms in \eq{Euler},
 we can also solve for the flow structure of the atmosphere as function of radius. Vortensity conservation (which replaces the continuum equation) and force balance give:
\begin{align}
  \label{eq:vort-simple}
  \frac{\partial}{\partial r} (rv_\phi)
  &= r \left(\frac{1}{2}\rho -2 \right) \\
  \frac{1}{\rho}\frac{\partial \rho}{\partial r}
  &= \frac{v_\phi^2}{r} -\frac{\partial\Phi}{\partial r}
  \label{eq:Euler-simple}
\end{align}
We solve this system numerically for $r_\mathrm{inn}\le r \le m$, for which circular motion is indeed reasonable.  This system of two equations requires two boundary conditions. It is natural to assume a hydrostatic profile at $r=m$, which fixes $\rho$ at $\exp(1)$. Let us further fix $v_\phi=0$ at the inner radius $r_\mathrm{inn}$, consistent with Kelvin's circulation theorem.

\Fg{atmos-profile} shows the atmosphere profiles from the numerical simulations (circles). The upper panel of \fg{atmos-profile} shows the azimuthal velocities and the lower panel the densities. The latter all overlap when $r\rightarrow m$. The solid curves give the predicted profile after numerical integration of \eqs{vort-simple}{Euler-simple}. The fits to the data are very good, especially for the density. For the azimuthal velocities curves (top panel) $v_\phi(r_\mathrm{inn})$ is significantly lower than the data when $r\rightarrow r_\mathrm{inn}$, which we attributed to numerical viscosity effects near the boundary. 

What we see in \fg{atmos-profile} is the transition from a pressure-supported disc to a rotationally-supported disc. Plotted in \fg{atmos-profile}a is also the Keplerian rotational velocity (dotted line\ch{: $v_\mathrm{Kepl}=\sqrt{m/r}$) the circular velocity corresponding to the adopted 2-body potential (\ie\ $v_\mathrm{circ}=\sqrt{r(\partial\Phi_\mathrm{2b}/\partial r)}$; dashed line)} and in \fg{atmos-profile}b the hydrostatic solution of the atmosphere $\rho_\mathrm{hs}$ (\eq{rho-hs}; dashed). For $r_\mathrm{inn}=0.2m$ and $0.1m$ the data are consistent with a hydrostatic solution (the $r_\mathrm{inn}=0.1m$ curve can hardly be distinguished). However for the smaller $r_\mathrm{inn}$ the hydrostatic solution no longer applies: velocities have now approached the circular velocity and rotational support enters the force balance.

\section{Discussion}
\label{sec:discuss}
\subsection{Steady state and geometry}
In this work we have conducted 2D inviscid, isothermal hydrodynamical simulations of the flow past low mass embedded planets, with a focus on the atmosphere of the planet -- the region where the gas starts to accumulate. For such low-mass planets (our standard planet mass of $m=10^{-2}$ corresponds to a Mars-size body at 1 AU; \fg{dimQs}) the gas in the atmosphere builds up slowly, on timescales of millions of years \citep{PollackEtal1996}, much larger than the dynamical timescales the hydrodynamical simulation can resolve. 

The prime aim of this study was to answer the question whether the flow is in steady state. To facilitate this task we adopted a polar geometry, with grid points logarithmically spaced in radius, so that the region between the inner boundary is guaranteed to be well resolved. We found that a polar grid greatly sped up the convergence towards a steady flow. Indeed, we found that for any combination of parameters (planet mass, inner radius, headwind) results in a steady flow, provided one ensures a sufficiently high resolution and a gradual injection of the 2-body potential.

In contrast the uniform Cartesian geometry simulation have much more difficulty to converge to steady solutions. We argued that it is not merely a matter of resolution as our highest Cartesian resolution runs have a comparable number of grid points within the Bondi radius than the polar runs. The polar geometry is advantageous because the flow becomes circular when $r\rightarrow0$ and because the polar runs do not include the problematic origin. Nevertheless, the Cartesian runs do show clear convergence behavior.  Adaptive or nested grid techniques are thus a preferred choice for Cartesian geometries \citep[\eg][]{D'AngeloBodenheimer2013}. 

Our findings are broadly in line with the literature, although these are often far more sophisticated in terms of the thermodynamic treatment, apply to higher-mass planets, and are conducted in 3D. \citet{BateEtal2003}, conducting simulations in 2D and 3D, find a quasi-steady state. Yet, their simulations apply a sink cell approach, which we argue is not appropriate for the low-mass regime. A similar treatment is performed by \citet{D'AngeloEtal2002}. \citet{D'AngeloEtal2003i,D'AngeloEtal2003} perform simulations without accretion, finding, like us, that the atmospheres of non-accreting planets are pressure supported. \citet{AyliffeBate2009} experiment with a `core' in their SPH simulations, very similar to the reflective boundary of our polar grid, include a realistic thermodynamic treatment, and find that accretion rates in their lowest mass simulation (10 Earth masses at 5.2 AU) are low. However, as these are still rather massive planets the contraction of the gaseous envelope due to cooling is not negligible; indeed, \citet{AyliffeBate2012} do observe that one of these atmospheres eventually collapses. Our key result of a steady atmosphere reflects \citet{D'AngeloBodenheimer2013}, despite the very detailed treatment of the thermodynamics in that study (even accounting for planetesimal heating). In contrast, \citet{NelsonRuffert2013} retrieve a very dynamic (unstable) atmosphere for a 10 $\mEarth$ planet, which they attribute to their non-isothermal equation of state. 

A proper thermodynamic treatment, as considered by many of the works above, is arguably more realistic than our simple isothermal EOS. In this paper, we opted for a simple EOS to isolate the thermodynamic evolution from the basic hydrodynamic properties of the flow \citep[\cf][]{MorbidelliEtal2014}. We believe that for low-mass planets, where evolutionary timescales are long, such a treatment is especially appropriate. Nevertheless, an obvious follow-up step is to replace the isothermal assumption by an adiabatic EOS or by radiation transport to verify the steady solution found here.

\subsection{Importance of rotation}
Having confirmed that the atmosphere around a low mass protoplanet is stable, we investigated what sets its size. By varying the inner radius $r_\mathrm{inn}$ we found that the atmosphere size increased, a consequence of Kelvin's circulation theorem. The amount of gas that is confined in the atmosphere ($M_\mathrm{atm}^{<r}$) therefore determines its size.  This is an interesting finding, because it implies that the global properties of the flow, the size of the atmosphere and the size of the horseshoe region, are determined by the physics of the atmosphere.  

\begin{figure}
  \centering
  \includegraphics[width=88mm]{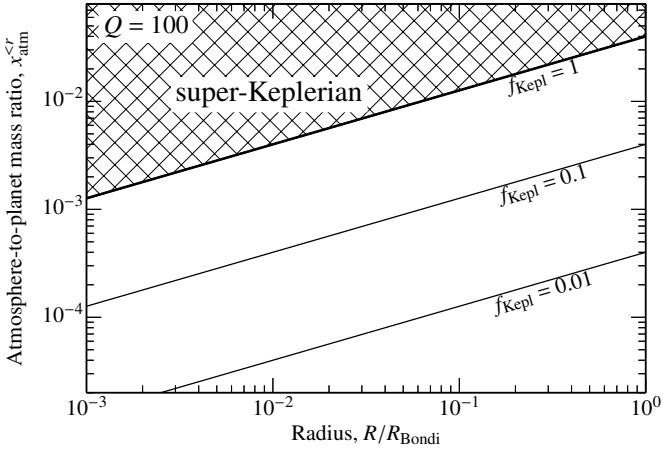}
  \caption{Contour plot of the rotational velocity as a fraction of Keplerian ($f_\mathrm{Kep}$) as function of radius and the cumulative gas mass fraction of the atmosphere, $x_\mathrm{atm}^{<r}=M_\mathrm{atm}^{<r}/M_p$ for a Toomre-Q parameter of 100. Discarding the super-Keplerian limit, this plot shows that atmosphere mass fractions can never exceed 5\% as gas becomes rotationally supported. The squares further constrain this limit by taking into account the requirement that the density must match the disc density at the Bondi radius. \ch{The lines shift upwards when the mass of the circumstellar disc increases (\ie\ proportional to $Q_T^{-1}$).}}
  \label{fig:rotation}
\end{figure}
\comno{The derivation below is a bit tricky as I had to take care about conversion to physical units. (The gas mass units is different than the planet mass!) Perhaps someone better checks these equations. Also instead of densities, we really have surface densities in 2D. I could ignore this subtlety but here it matters.}
We can generalise the results of \se{rinn} to planets and atmospheres of arbitrary masses.  \Eq{Kelvin-sol} gives, in dimensionless units, the azimuthal velocity as function of radius. \ch{For sufficiently small $r$ the $r$-term in \eq{Kelvin-sol} (read: $r\Omega$) becomes insignificant compared to $v_\phi$ (see \fg{atmos-profile}). Therefore, we can write} $v_\phi \approx \mathscr{M}_\mathrm{atm}^{<r}/4\pi r$. Expressed in terms of the Keplerian velocity $v_\mathrm{Kepl}=\sqrt{m/r}$, this reads $f_\mathrm{Kepl} \equiv v_\phi/v_\mathrm{Kepl} = \mathscr{M}_\mathrm{atm}^{<r}/4\pi \sqrt{mr}$.

It is more useful to express $f_\mathrm{Kepl}$ in terms of the atmosphere-to-planet mass fraction $x_\mathrm{atm}^{<r}$. Reminding the reader that the dimensionless \textit{gas} mass $\mathscr{M}_\mathrm{atm}^{<r}$ is in units of $\Sigma_\mathrm{disc} H^2$ while that of the planet's \textit{gravitating} mass is in units of $c_s^3/G\Omega$ (see \se{dimensions}), the relation between $x_\mathrm{atm}$ and the dimensionless atmosphere mass is:
\begin{equation}
  x_\mathrm{atm}^{<r}
  = \frac{M_\mathrm{atm}^{<r}}{M_p}
  = \frac{ \mathscr{M}_\mathrm{atm}^{<r} \Sigma_\mathrm{disc} H^2 }{ m c_s^3 /G\Omega}
  = \frac{ \mathscr{M}_\mathrm{atm}^{<r} }{\pi m Q_T},
\end{equation}
where $Q_T=c_s \Omega/\pi G\Sigma_\mathrm{disc}$ is Toomre's-Q of the circumstellar disc. Using this relation to eliminate $\mathscr{M}_\mathrm{atm}^{<r}$ we express the Keplerian fraction in terms of the atmosphere-to-planet mass:
\begin{equation}
  f_\mathrm{Kepl}
  = \frac{x_\mathrm{atm}^{<r} Q_T}{4} \sqrt{\frac{m}{r}}
  = \frac{x_\mathrm{atm}^{<r} Q_T}{4} \sqrt{\frac{R_\mathrm{Bondi}}{R}}
  \label{eq:fKeplII}
\end{equation}
(where the last expression gives the dimensional form).  This is illustrated in \fg{rotation}, where we plot contours of $f_\mathrm{Kepl}$ as function of the atmosphere radius $r$ (x-axis) and the atmosphere mass fraction $x_\mathrm{atm}^{<r}$ (y-axis) for a Toomre parameter of $Q_T=100$. Thus, for a given atmosphere mass (horizontal line) $f_\mathrm{Kep}$ decreases with radius. For a given radius (vertical line) $f_\mathrm{Kep}$ increases with increasing $x_\mathrm{atm}^{<r}$.

The $f_\mathrm{Kep}=1$ line puts an upper limit to the atmosphere mass. \ch{For $Q_T=100$ this means that the atmosphere will become rotationally supported long before it reaches the critical mass ($x_\mathrm{atm}^{<r}\sim1$). However, the lines in \fg{rotation} shift proportional to disc density; for lower $Q_T$-values (more massive discs) rotationally support is thus much less a concern. This holds in particular when discs are massive enough to self-gravitate ($Q_T\sim1$; \citealt{MayerEtal2004}).} 

\ch{In the case of Keplerian rotation ($f_\mathrm{Kepl}=1$) the atmosphere mass reads $\mathscr{M}_\mathrm{atm}^{<r}=4\pi\sqrt{mr}$; and the corresponding gas density is}
\begin{align}
    \nonumber
    \Sigma_\mathrm{Kepl}
    &= \frac{1}{2\pi r} \frac{d}{dr} \left( \mathscr{M}_\mathrm{atm}^{<r} \right)_\mathrm{f_\mathrm{Kepl}=1}
    = \frac{1}{m} \left( \frac{m}{r} \right)^{3/2} \\
    &= \Sigma_\mathrm{disc} h^3 \left( \frac{M_p}{M_\star} \right)^{-1} \left( \frac{R_\mathrm{Bondi}}{R} \right)^{3/2}
    \label{eq:sigma}
\end{align}
(where the bottom line gives the physical units with $h$ the disc aspect ratio). \Eq{sigma} is the density profile of a Keplerian-rotating \textit{circumplanetary} disc for low planet masses. There is only a single solution for $\Sigma_\mathrm{Kepl}$ by virtue of the constraint that it has the same vortensity as the circumstellar disc material. It will thus have a $\Sigma\propto R^{-3/2}$ density profile.

\combox{I still struggle with the link to the 3D results. We do not want to be too bold here, but also not that these conclusions are academic. And then there is the Wang study..}
We emphasise the generality of the results as shown in \fg{rotation}. 
\ch{These follow from basic angular momentum considerations (vortensity conservation): the gas that makes up the atmosphere originated from a Keplerian-rotating circumstellar disc which, in the absence of shocks, conserves its vortensity. This holds irrespective whether the EOS is isothermal or not. The inviscid assumption, which also enters, is of course subject\? to any viable mechanism for angular momentum transport operating in such discs. Assessing their stability is beyond the scope of this work, but we note that the for circumstellar discs frequently invoked magneto-rotational instability is unlikely to be very effective \citep{FujiiEtal2014}.} 

\ch{Perhaps the biggest caveat is the 2D assumption (vortensity conservation); it is thus to be seen whether 3D calculations of low-mass planets will give a similar result. However, recent 3D global simulations by \citet{WangEtal2014} find (near-)Keplerian rotating circumplanetary discs around low-mass planets, which lends some support to the idea of a rotational bottleneck for atmosphere growth, described above.}

\subsection{Outlook}
In this work we have employed idealised assumptions as 2D and inviscid and focused on the low-mass regime. The advantage of these are that we can use the conservation of vortensity to keep track of the fidelity of our simulation and that we could present a parameter study as 2D calculations are relatively cheap. 

\ch{We have assumed that the flow is inviscid, which preserves the vortensity. For a viscous flow, angular momentum can be transported outwards which would ameliorate the rotation bottleneck described above. 
Conversely, angular momentum may also be transported inwards and/or deposited on the core, causing it to spin up. Simulations that include viscosity are needed to describe how efficient the spinning-up mechanism operates.}

\ch{As our focus lies on low mass, embedded planets, a natural step is to conduct the simulations in 3D using a spherical coordinate system. }
A disadvantage here is that the spherical grid contains another singularity: \ch{near the poles, grid points are spaced at distance $\sim$$2\pi r\theta/N_\phi$ with $N_\phi$ the number of azimuthal grid points. As this becomes very small for $\theta\rightarrow0$ the Courant condition results in a timestep that is likewise short.} Calculations are thus expected to be much more time-consuming. In addition, in 3D the vortensity will no longer be conserved\com{But what about Kelvin's theorem? I think it will..}. In a successor work we will re-assess the questions raised in this paper---whether the atmosphere flow is in steady state and whether it is bound---but for a 3D geometry.

Finally, we can extend our work to higher-mass planets. The gas near planets with $R_\mathrm{Bondi}>H$ ($m>1$) is better approximated as 2D because the scaleheight is small compared to the Bondi radius. Super-Earth planets ($\sim$10 $\mEarth$ at 0.1 AU) fall into this category as they have $m\sim3$ (see \fg{dimQs}). For $m>1$ planets, however, the spiral density perturbation, which is small for low-mass planets, will turn nonlinear \citep{LinPapaloizou1993}, which generates vortensity and results in a disc in which vortensity is no longer conserved \citep{KollerEtal2003,LiEtal2005}. We will investigate the properties of such discs and thereby revisit the question whether rotation will limit the growth of the atmosphere as we see in \fg{rotation}. An addition area of exploration is the consequence of a viscosity on the evolution of such discs.
\com{more ideas?\dots}

\section{Summary}
\label{sec:summary}
\combox{Yeah, a summary! Do read!}
\ch{In this work, we have investigated the properties of an inviscid and isothermal flow past a low mass planet, embedded in a gas-rich circumstellar disc. We employed the \pluto\ hydrodynamic code using a 2D polar or Cartesian grid with the planet at its center. Gas is not removed from the simulation domain (no sink cell) and smoothing of the Newtonian potential has been considerably suppressed to resolve scales much smaller than the Bondi sphere. Our key findings are:
\begin{enumerate}
    \item The 2D flow is steady. Circulating streamlines demarcate the bound atmosphere material from the disc. The shape and size of the atmosphere depend strongly on the flow pattern of the background. In particular, the headwind arising from a sub-Keplerian rotating circumstellar disc strongly affects the flow pattern. 
    \item The polar geometry features a logarithmic grid spacing and an inner boundary, which greatly benefit the fidelity of the simulation, giving accurate results already at low resolution. In contrast, vortensity conservation is problematic in Cartesian grids. These must be conducted at sufficiently high resolution to prevent the flow from becoming unsteady.
    \item Decreasing the inner radius of the domain $r_\mathrm{inn}$, causes the atmosphere to compress and accrete more gas and to increase its size. In agreement with Kelvin's circulation theorem, we saw a correlation between the atmosphere azimuthal velocity and the enclosed mass. 
    \item This implies that atmospheres of low-mass planets will reach Keplerian velocities, and therefore rotational support, before their atmospheric mass fractions  become comparable to the gas. For large Toomre-Q values of the circumstellar disc, this implies a centrifugal barrier against the collapse of the circumplanetary atmosphere.
\end{enumerate}
The latter is a strong conclusion subjected to the idealizations employed in this work, of which we believe the 2D is the strongest. Nevertheless, this work has demonstrated that angular momentum considerations enter the picture on the formation of low mass planets, as it does in the formation theories of high-mass planets, stars, and galaxies.}

\section*{Acknowledgments}
This work has profited immensely from discussion with Gennaro D'Angelo, Eugene Chiang, Kees Dullemond, Hubert Klahr, Willy Kley, Andrea Mignone, John Ramsey, Neal Turner, and other colleagues.  CWO would like to thank Thomas Henning for his suggestion to use the \pluto\ code, which inspired this project. For CWO support for this work was provided by NASA through Hubble Fellowship grant \#HST-HF-51294.01-A awarded by the Space Telescope Science Institute, which is operated by the Association of Universities for Research in Astronomy, Inc., for NASA, under contract NAS 5-26555. RK acknowledges funding from the Max Planck Research Group `Star formation throughout the Milky Way Galaxy' at the Max Planck Institute for Astronomy.

\bibliographystyle{apj}
\bibliography{ads,arXiv}

\appendix
\section[]{Non-gravitational flow past a sphere}
\label{app:tests}
\label{app:sphere}
\begin{figure}
  \includegraphics[width=88mm]{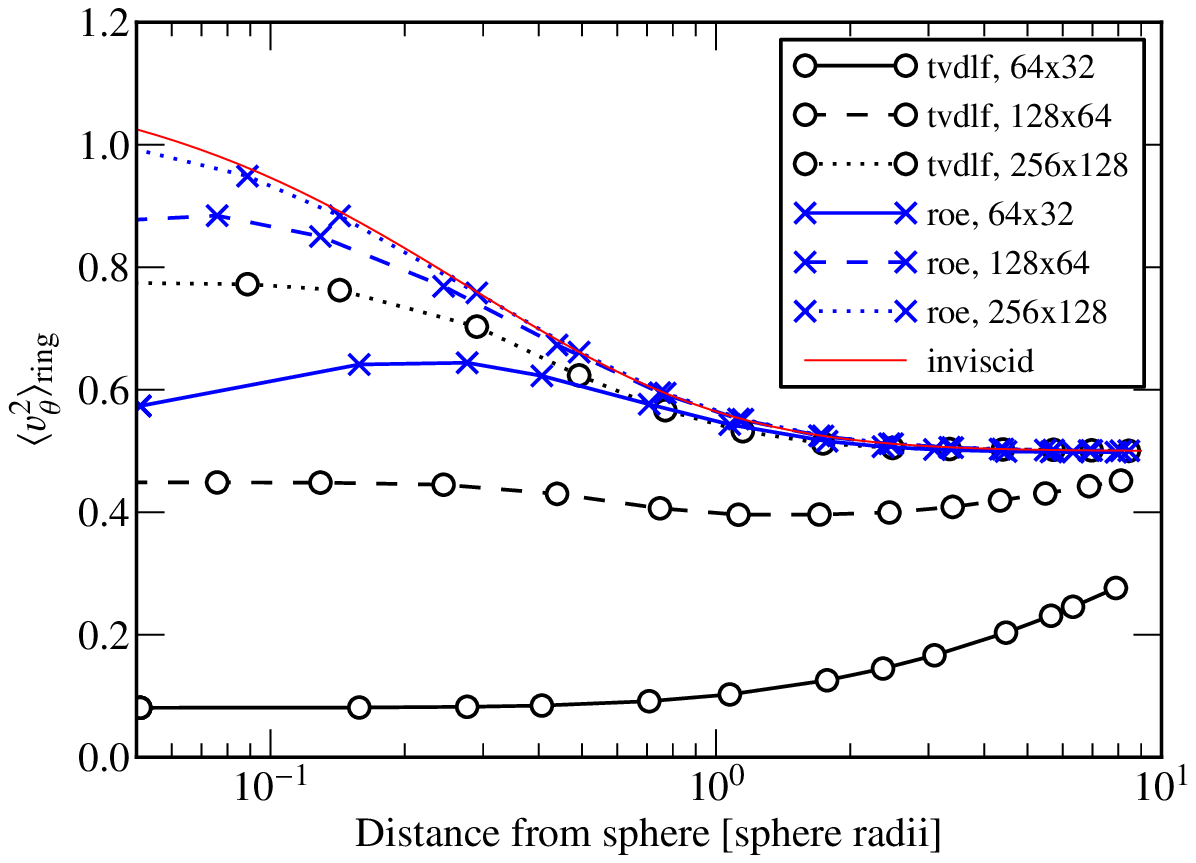}
  \caption{Estimate of the effect of numerical viscosity. Shown is the azimuthally-averaged square of the azimuthal component of the velocity $v_\theta^2$ as function of distance from the sphere. The analytical inviscid solution is shown by the upper thin red solid curve. It increases with decreasing distance to the sphere. The numerical solutions are shown for the \texttt{tvdlf} solver (black, spheres) and the \texttt{roe} solver (blue, crosses) at three different resolutions (linestyle). Numerical viscosity results in a smaller velocity. }
  \label{fig:sphere-test}
\end{figure}
The inviscid solution for irrotational, incompressible flow past a sphere reads \citep[\eg][]{LandauLifshitz1959}:
\begin{align}
  \label{eq:v-inviscid-rad}
    v_r       &=  \mathcal{M}_\mathrm{hw} \cos\theta \left( 1 -\frac{a^3}{r^3} \right) \\
    v_\theta  &= -\mathcal{M}_\mathrm{hw} \sin\theta \left( 1 +\frac{a^3}{2r^3} \right)
  \label{eq:v-inviscid-theta}
\end{align}
where ${\cal M}$ is the velocity of the unperturbed flow, $a$ the radius of the sphere, $r\ge a$ the distance from the center of the sphere and $\theta$ the polar angle.  The incident flow is directed towards the $Z$-axis ($\theta=0$).

In \fg{sphere-test} we have tested how well \pluto\ matches this solution for various choices of the resolution and solver options. By employing the reflective boundary conditions (as in the main paper) we expect the irrotational solution of \eqs{v-inviscid-rad}{v-inviscid-theta} to hold, in contrast to the real flow solutions (which feature no slip boundary conditions). To estimate the amount of numerical viscosity we average the square of the azimuthal component of the velocity, $v_\theta^2$ over a ring at radius $r$. Thus, for the inviscid solution of \eq{v-inviscid-theta} $\langle v_\theta^2 \rangle$ runs from $9\mathcal{M}_\mathrm{hw}^2/8$ at the sphere radius ($r=a$) to ${\cal M}^2/2$ as $r \gg a$ as shown by the red thin curve in \fg{sphere-test}.

The other lines give the numerical results. The initial state is the unperturbed solution: uniform flow moving along the polar axis (the $Z$-axis). The geometry is spherical with azimuthal symmetry with the radial grid points logarithmically distributed. The radius of the sphere is $r_\mathrm{inn}=10^{-3}$, and the outer radius of the domain is $r=0.5$. The inner boundary condition is chosen as reflective and the outer obeys the unperturbed flow.  This reflects the conditions of the main paper (but without shear and body forces). The incoming headwind is chosen low enough, $\mathcal{M}_\mathrm{hw}=10^{-2}$, to discount compressibility effects affecting the analytical solution.

The results are shown in \fg{sphere-test} for the \texttt{tvdlf} and the \texttt{roe} solvers \com{Explain here or in the main chapter what these solvers are and give references.} for three different resolutions: low (64 radial grid points), medium (128) and high (256). The deviation from the inviscid curve thus give a measure for the amount of numerical viscosity. Clearly, this is a function of numerical resolution with finer grids having less viscosity. (Note that we have intentionally chosen a very coarse grid.) It also depends on the choice for the algorithm to solve the hydrodynamical equations, with the roe algorithm, which solves the Riemann problem, producing a far better result than the \texttt{tvdlf} algorithm\refs. In fact, at the lower resolution the velocity for the \texttt{tvdlf} solver drops below that of the unperturbed flow ($\mathcal{M}_\mathrm{hw}/2$)---a situation more reminiscent to flow in the Stokes regime where $v_\theta(r=a)=0$. 

\bsp
\label{lastpage}
\end{document}